\documentclass[a4paper,usenames,dvipsnames,11pt]{article}

\usepackage{jheppub}
\usepackage{slashed}
\usepackage{mathrsfs,booktabs}
\usepackage{stmaryrd}
\usepackage{xspace}
\usepackage{fancyvrb}
\usepackage[makeroom]{cancel}

\usepackage[normalem]{ulem}

\def\showcommentsflag{0}
\newcommand{\showcomments}{\def\showcommentsflag{1}}

\newcounter{commentcounter}%
\newcommand{\comment}[1]{\ifnum\showcommentsflag > 0%
\addtocounter{commentcounter}{1}%
\Red{\ensuremath{\ddagger^{\arabic{commentcounter}}}}%
\marginpar{\raggedright\tiny\it\Red{\ensuremath{\ddagger^{\arabic{commentcounter}}} #1}}
\fi%
}
\newcommand{\commentdel}[2]{\ifnum\showcommentsflag > 0%
\Red{\sout{#1}}\comment{#2}%
\fi
}
\newcommand{\commentadd}[2]{\ifnum\showcommentsflag > 0%
\comment{#2}\Red{#1}%
\else
#1
\fi
}
\newcommand{\commentchange}[3]{\ifnum\showcommentsflag > 0%
\Red{\sout{#2}}\comment{#3}\Red{#1}%
\else
#1
\fi
}
\newcommand{\nocomment}[1]{\ifnum\showcommentsflag > 0%
{\tiny\it\Red{\{#1}\}}
\fi%
}
\newcommand{\nocommentdel}[1]{\ifnum\showcommentsflag > 0%
\Red{\sout{#1}}%
\fi
}
\newcommand{\nocommentadd}[1]{\ifnum\showcommentsflag > 0%
\Red{#1}%
\else
#1
\fi
}
\newcommand{\nocommentchange}[2]{\ifnum\showcommentsflag > 0%
\Red{\sout{#2}}\Red{#1}%
\else
#1
\fi
}

\showcomments

\definecolor{myred}{rgb}{0.9,0.05,0.05}
\newcommand{\Red}{\textcolor{myred}}

\newcommand{\eq}[1]{eq.~\eqref{eq:#1}}
\newcommand{\eqs}[2]{eqs.~\eqref{eq:#1} and \eqref{eq:#2}}
\renewcommand{\sec}[1]{section~\ref{sec:#1}}

\newcommand{\subsec}[1]{section~\ref{subsec:#1}}
\newcommand{\subsecs}[2]{sections~\ref{subsec:#1} and \ref{subsec:#2}}
\newcommand{\app}[1]{Appendix~\ref{app:#1}}
\newcommand{\fig}[1]{figure~\ref{fig:#1}}

\newcommand{\mycite}[1]{ref.~\cite{#1}}
\newcommand{\mycites}[1]{refs.~\cite{#1}}
\newcommand{\tab}[1]{table~\ref{tab:#1}}

\newcommand{\as}{\alpha_s}

\allowdisplaybreaks

\newcommand{\herwig}{{\sc{Herwig6}}\xspace}
\newcommand{\pythia}{{\sc{Pythia8}}\xspace}
\newcommand{\mga}{{\sc{MadGraph5\_aMC@NLO}}\xspace}
\newcommand{\ms}{{\sc{MadSpin}}\xspace}
\newcommand{\ml}{{\sc{MadLoop}}\xspace}
\newcommand{\ct}{{\sc{CutTools}}\xspace}
\newcommand{\mf}{{\sc{MadFKS}}\xspace}
\newcommand{\mcnlo}{{\sc{MC@NLO}}\xspace}

\newcommand{\bea}{\begin{eqnarray}}
\newcommand{\eea}{\end{eqnarray}}
\newcommand{\xcut}{x_{\text{cut}}}
\newcommand{\mtrec}{m_t^{\text{rec}}}
\newcommand{\mbrec}{m_\beta^{\text{rec}}}
\newcommand{\bwcut}{{\texttt{BWcut}}}
\newcommand\prompt{{\tt MG5\_aMC>}}

\numberwithin{equation}{section}

\def\beq{\begin{equation}}
\def\beqn{\begin{eqnarray}}
\def\eeq{\end{equation}}
\def\eeqn{\end{eqnarray}}

\newcommand\sss{\scriptscriptstyle}

\newcommand\kbt{k_\beta^2}
\newcommand\kbtmo{k_\beta^{2^{-1}}}
\newcommand\mb{m_\beta}
\newcommand\mbt{m_\beta^2}
\newcommand\Gb{\Gamma_\beta}

\newcommand\boost{\mathbb B}
\newcommand\xih{\hat{\xi}}
\newcommand\ximax{\xi^{\sss\rm M}}
\newcommand\stepf{\Theta}

\newcommand\bbe{b_\beta}

\newcommand\bnotbe{b_{\xcancel{\beta}}}

\newcommand\sep{\,|\,}
\newcommand\phivar{b}
\newcommand\pt{p_{\sss T}}

\VerbatimFootnotes

\title{\bf Off-shell single-top production
at NLO matched to parton showers}

\author[a]{R. Frederix,}
\author[b]{S. Frixione,}
\author[c]{A.S. Papanastasiou,}
\author[d]{S. Prestel,}
\author[e]{and P. Torrielli}
\affiliation[a]{Physik Department T31, Technische Universit\"at M\"unchen, 
James-Franck-Str. 1,\\ D-85748 Garching, Germany}
\affiliation[b]{INFN, Sezione di Genova, Via Dodecaneso 33, I-16146, 
Genoa, Italy}
\affiliation[c]{Cavendish Laboratory, University of Cambridge, 
J.J. Thomson Avenue,\\ CB3 0HE, Cambridge, UK}
\affiliation[d]{SLAC National Accelerator Laboratory, 2575 Sand Hill Road,\\ 
Menlo Park, CA 94025-7090 USA}
\affiliation[e]{Dipartimento di Fisica, Universit\`a di Torino and INFN,
 Sezione di Torino,\\ Via P.~Giuria~1, I-10125, Turin, Italy}

\emailAdd{rikkert.frederix@tum.de}
\emailAdd{Stefano.Frixione@cern.ch}
\emailAdd{andrewp@hep.phy.cam.ac.uk}
\emailAdd{prestel@slac.standford.edu}
\emailAdd{torriell@to.infn.it}

\abstract{We study the hadroproduction of a $Wb$ pair in association with a
  light jet, focusing on the dominant $t$-channel contribution and including
  exactly at the matrix-element level all non-resonant and off-shell effects
  induced by the finite top-quark width.  Our simulations are accurate to the
  next-to-leading order in QCD, and are matched to the \herwig and \pythia
  parton showers through the \mcnlo method.  We present phenomenological
  results relevant to the 8 TeV LHC, and carry out a thorough comparison to
  the case of on-shell $t$-channel single-top production.  We formulate our
  approach so that it can be applied to the general case of matrix elements
  that feature coloured intermediate resonances and are matched to parton
  showers.}


\preprint{
\begin{flushright}
Cavendish-HEP-16/03\\
TUM-HEP-1037/16\\
\today
\end{flushright}
}

\begin{document}
\maketitle
\flushbottom

\newpage
\section{Introduction}
Single-top production at hadron colliders has continued to be an active field
of research, challenging both experimental and theoretical communities,
since its observation at the Tevatron~\cite{Aaltonen:2009jj,Abazov:2009ii}.
Measurements of the dominant $t$-channel subprocess have been presented 
by CDF and D0~\cite{Aaltonen:2010jr,Abazov:2011rz,Aaltonen:2015cra} 
at the Tevatron, as well as by the ATLAS~\cite{Aad:2012ux,Aad:2014fwa,
ATLAS:2012fpa} and CMS~\cite{Chatrchyan:2011vp,Chatrchyan:2012ep,
Khachatryan:2014iya} collaborations, at both the 7~TeV and 8~TeV LHC, with 
preliminary results~\cite{ATLAS-CONF-2015-079,CMS-PAS-TOP-15-004} 
available at 13~TeV as well.
This channel has also been exploited by the ATLAS collaboration in the 
first top-mass extraction from single-top events in 
\mycite{st-tch-mass}. More recently, experimental evidence has been 
found~\cite{Chatrchyan:2012zca,CDF:2014uma,Aad:2015eto,Aad:2015upn}
of $s$- and $Wt$-channel production, which are characterised by
cross sections smaller than that of the $t$ channel. Single top
production has been shown to be sensitive to anomalous $tWb$-couplings (see
for example~\mycites{Zhang:2010dr,Cao:2015doa}), and efforts are being made 
by experiments~\cite{CMS:2014ffa,Aad:2015yem} to use the $t$-channel 
process to search for such features. Furthermore, with increased statistics 
at run II of the LHC, measurements of the top-quark decay products and of 
differential quantities will be possible with vastly improved precision.

With a view to matching the progress achieved on the experimental side, it is
important to review, assess and improve the current theoretical predictions
available for single-top production.  Top quarks are never observed as stable
particles, but rather their production is inferred through a kinematic
reconstruction of their decay products (jets, leptons and missing energy).
Theoretical predictions, whenever possible, should therefore reflect this
fact, namely they should deal with top decay products instead of stable top
quarks as primary objects. This is particularly important for observables
sensitive to the decay and off-shellness of the top, as well as for those
sensitive to non-resonant contributions, which are completely missing in the
stable-top approximation.

Nonetheless, the current theoretical standard only partially fulfils this
requirement. State-of-the-art predictions at the hadron level for this process
are obtained through NLO-matching with parton showers (NLO+PS) both in the
four- and five-flavour schemes~\cite{Frixione:2005vw,
Alioli:2009je,Frederix:2012dh} in the \mcnlo~\cite{Frixione:2002ik} and 
{\sc POWHEG}~\cite{Nason:2004rx,Frixione:2007vw} approaches, 
assuming stable-top hard matrix elements. In such setups, the
top-quark decay is performed with LO accuracy, and the off-shellness of the 
top propagator is introduced through a simple Breit-Wigner smearing,
either by the PS itself, or at the matrix-element level (which allows
one to correctly account for both production and decay spin correlations)
by applying the method introduced in \mycite{Frixione:2007zp}. At fixed
order, alongside the NLO corrections to the production, NLO corrections to the
decay of the top quark have been included in the narrow-width approximation
(NWA)~\cite{Campbell:2004ch,Heim:2009ku,Schwienhorst:2010je,Campbell:2012uf}.
A systematic treatment of off-shell effects for resonant top quarks was first
presented in \mycites{Falgari:2010sf,Falgari:2011qa}, using an
effective-theory-inspired generalisation of the pole expansion.  The NLO
corrections to the $t$-channel process with full off-shell and non-resonant
effects have been computed in \mycite{Papanastasiou:2013dta} by adopting the
complex-mass scheme (CMS)~\cite{Denner:1999gp,Denner:2005fg}.

Including NLO QCD corrections to the top decay has been shown to play a
significant role, especially for observables such as transverse momentum of
the $b$-jet or the invariant mass of the lepton+$b$-jet 
system~\cite{Campbell:2012uf,Melnikov:2009dn}.  Additionally, 
treating the top quark as on-shell (as in the NWA) or off-shell can also lead to
striking differences in the NLO predictions of experimentally 
relevant observables~\cite{Falgari:2010sf,Falgari:2011qa,
Papanastasiou:2013dta}, a prime example being the invariant mass of the
reconstructed top (see also
\mycites{Denner:2010jp,AlcarazMaestre:2012vp,Falgari:2013gwa} for similar
features in $t\bar{t}$ production).  In light of these observations at fixed
order, understanding to what extent these effects survive the showering and
hadronisation stages in a Monte Carlo (MC) is not only interesting from the
theory point of view, but it also becomes crucial for improved predictions of
the observables mentioned above. In particular, with predictions at NLO+PS
accuracy and full off-shell effects at the hard matrix-element level, it
becomes possible to validate NLO+PS approaches where the underlying hard
matrix elements are computed in the on-shell-top approximation.  It is also of
great relevance to use these improved predictions to properly assess the
systematics affecting the extraction of the top mass when using, as 
is currently done, MCs within which the hard matrix elements do not include 
full NLO off-shell effects.  Recently, work has been performed in this
direction in \mycites{Campbell:2014kua,Jezo:2015aia} within the 
{\sc POWHEG}$+$\pythia\ framework, including NLO corrections in both
production and decay, and considering $t\bar{t}$ production in the NWA and
single-top $t$-channel production with full off-shell effects, respectively.

In this work we adopt the \mcnlo scheme, and study the NLO matching to 
parton showers of $t$-channel single-top hadroproduction with full off-shell 
and non-resonant effects, namely the $t$-channel contribution to the EW process
$pp\to W^+bj$, with $j$ being a light jet, at the 8 TeV LHC.  We match our 
computations to the \herwig~\cite{Corcella:2000bw,Corcella:2002jc} and 
\pythia~\cite{Sjostrand:2007gs} parton showers.  
In this context, we discuss in general the subtleties that occur in NLO+PS
simulations for processes with intermediate coloured resonances, and perform 
a thorough comparison to other available approximations of $t$-channel 
single-top cross section.  In doing so, we present a study of hadron-level 
observables sensitive to top-decay radiative corrections and off-shell effects.
The shape of such observables is often a result of a sensitive interplay of a
number of different phenomena, which we endeavour to disentangle and understand
here.

We perform our calculations in the framework of \mga~\cite{Alwall:2014hca},
which automates all ingredients relevant to the simulation of LO and NLO cross
sections, including the matching to parton showers.  
The FKS method~\cite{Frixione:1995ms,Frixione:1997np} (automated in
the module \mf~\cite{Frederix:2009yq}) for the subtraction
of the infrared (IR) singularities of real-emission matrix elements
underpins all NLO-accurate results. The computations of one-loop amplitudes 
are carried out by switching dynamically between two integral-reduction
techniques, OPP~\cite{Ossola:2006us} and TIR~\cite{Passarino:1978jh,
Davydychev:1991va,Denner:2005nn}. These have been automated in the 
module \ml~\cite{Hirschi:2011pa}, which in turn exploits 
\ct~\cite{Ossola:2007ax} together with an in-house implementation of the 
{\sc OpenLoops} optimisation~\cite{Cascioli:2011va}. Matching to parton 
showers is achieved by means of the \mcnlo formalism~\cite{Frixione:2002ik}.

The paper is structured as follows: in \sec{setup} we describe the setup of
the computation, and in particular the subtleties related to the phase-space
parametrisation, integration, and MC@NLO-type matching of processes with 
intermediate coloured resonances; in \sec{results} we discuss some details
of the various approximations to the complete $W^+bj$ process, and we present 
our results for a selected set of observables; in \sec{conclusion} 
we draw our conclusions.

\section{Matching setup, subtleties and technicalities\label{sec:setup}}

\subsection{Integration of subtracted cross sections with resonances
\label{sec:gen}}
Regardless of whether a computation that features an
unstable intermediate particle (that henceforth we denote by $\beta$, 
and assume to have mass $\mb$ and width $\Gb\ll\mb$) is matched to parton 
showers, one problem which must be addressed is that of the efficient 
integration of the corresponding matrix elements. This is particularly 
non-trivial in the case where these matrix elements enter the real-emission 
contribution to an NLO cross section, owing to the necessity of 
IR-subtracting them. At the amplitude level, the unstable particle
is represented by an $s$-channel propagator and thus the matrix
elements will contain a term
\beq
\frac{1}{(\kbt-\mbt)^2+(\Gb\mb)^2}\,,
\label{eq:betaprop}
\eeq
with $\kbt>0$ the virtuality of $\beta$. 
Because of \eq{betaprop}, kinematic configurations 
with $\kbt\simeq\mbt$ will be associated with large weights, and hence 
the corresponding unweighted events will be more likely
to occur. The likelihood of this increases with decreasing $\Gb$, which 
is easy to understand also in view of the fact that the $\Gb\to 0$ limit 
of \eq{betaprop} is proportional to the Dirac delta function 
$\delta(\kbt-\mbt)$, that forces $\kbt=\mbt$ exactly.
An efficient matrix element integration therefore requires that the 
phase-space generation be biased towards $\kbt\simeq\mbt$ configurations,
a requirement which is independent of the perturbative order.
At the LO (i.e.~tree) level, this is not difficult to achieve.
The most direct way is that of choosing $\kbt$ as one of the integration 
variables, so that the adaptive integration quickly knows where
to throw most of the phase-space points. This is what is done in \mga. 
While one would like to apply a similar strategy at the NLO and beyond, 
it is the IR subtractions relevant to the real-emission terms that prevent 
one from doing this in a straightforward manner. In contrast, all of the other
non-subtracted contributions to an NLO cross section, such as the Born 
and the virtuals, can be dealt with in exactly the same way as the 
LO. In order to simplify the discussion of the relevant issues
without loss of generality, let us assume that only one type of singularity 
is relevant (say, the soft singularities). Following FKS, we shall denote by 
$\xi$ the phase-space variable that in the limit $\xi\to 0$ causes the matrix 
elements to be soft-singular, and by $\phivar$ all of the other (Born-level)
phase-space  variables. The typical structure of the integrated NLO cross 
section will thus be:
\beq
\int d\phivar d\xi \frac{1}{\xi}\left[
\varsigma\!\left(\kbt(\phivar,\xi)\sep\phivar,\xi\right)-
\varsigma\!\left(\kbt(\phivar,0)\sep\phivar,0\right)\right]\,,
\label{eq:sig1}
\eeq
where the redundant first argument ($\kbt$) of the integrand $\varsigma$ 
($\varsigma$ is equal to the matrix elements times phase-space factors) has 
been inserted explicitly only in view of its relevance to the present 
discussion. Because of the way \eq{sig1} is integrated (i.e.~by choosing
some $\phivar$ and $\xi$ for any given random number), its event and 
counterevent contributions (first and second term, respectively,
under the integral sign in \eq{sig1}) will typically have very 
different weights, owing to \eq{betaprop}, {\em unless} 
the condition
\beq
\kbt(\phivar,\xi)=\kbt(\phivar,0)\,,\;\;\;\;\;\;\;\;
\forall\,\xi\,,\;\;\;\phivar~{\rm given}
\label{eq:kbcond}
\eeq
is fulfilled. Such a difference in weights is responsible for a
poorly-convergent integration. 

In principle, this is simply an efficiency problem, since the convergence 
in the large-statistics limit is guaranteed by the condition
\beq
\lim_{\xi\to 0}\kbt(\phivar,\xi)=\kbt(\phivar,0)\,,
\label{eq:lim}
\eeq
which holds regardless of whether \eq{kbcond} is true or not.
In practice, however, the statistics one needs to accumulate rapidly grows 
with the inverse of $\Gb$, becoming infinite in the $\Gb\to 0$ limit.
Indeed, it is instructive to consider the situation in the limiting case
where \eq{betaprop} is replaced by a Dirac delta. When this
happens and the condition of \eq{kbcond} is not fulfilled,
then for any given $\phivar$ either the event or the counterevent
is non-null, but never both simultaneously (since it is either
$\kbt(\phivar,\xi)=\mbt$ or $\kbt(\phivar,0)=\mbt$). This implies that the
phase space is partitioned into two disjoint regions, in which either only 
the event or only the counterevent contribution to the integrand of 
\eq{sig1} is non-null, which in turn renders the numerical integration
impossible with finite statistics. Alternatively, and from a more physical
viewpoint, in processes where $\Gb$ is very small, for the majority of
phase-space points the event is computed at the resonance peak, while the
counterevents are far away from the peak (or vice versa), even though 
the energy of the emitted parton ($\xi$ in \eq{sig1}) might be small 
as well. Thus, a small-width resonance severely hampers the cancellation 
between event and counterevents. A viable solution stems
from a re-mapping of the phase space:
\beq
\phivar\;\longrightarrow\;\Phi_\xi(\phivar)\,.
\label{eq:phspmap}
\eeq
There is ample freedom in choosing the specific form of \eq{phspmap},
but nevertheless we can distinguish two classes of re-mappings.
The members of the first class fulfill the following condition:
\beq
\kbt(\phivar,\xi)=\kbt(\Phi_\xi(\phivar),0)\,,\;\;\;\;\;\;\;\;
\forall\,\xi\,,
\label{eq:kbcond2}
\eeq
for any given $\phivar$. This can be exploited by rewriting \eq{sig1} as 
follows:
\beq
\int d\phivar d\xi \frac{1}{\xi}\left[
\varsigma\!\left(\kbt(\phivar,\xi)\sep\phivar,\xi\right)-
\frac{\partial\Phi_\xi(\phivar)}{\partial\phivar}\,
\varsigma\!\left(\kbt(\Phi_\xi(\phivar),0)\sep\Phi_\xi(\phivar),0\right)
\right]\,.
\label{eq:sig2}
\eeq
In other words, \eq{phspmap} is used for changing the
integration variables of the counterevent (see \app{reso} for a 
few technical details on this procedure). Thanks to this, the virtuality of 
$\beta$ is the same in the event and counterevent of \eq{sig2}, hence solving 
the original problem. Conversely, the re-mappings that belong to the
second class fulfill the condition:
\beq
\kbt(\Phi_\xi(\phivar),\xi)=\kbt(\phivar,0)\,,\;\;\;\;\;\;\;\;
\forall\,\xi\,,
\label{eq:kbcond3}
\eeq
for any given $\phivar$. 
One thus changes the integration variables of the event contribution,
whence the analogous of \eq{sig2} reads:
\beq
\int d\phivar d\xi \frac{1}{\xi}\left[
\frac{\partial\Phi_\xi(\phivar)}{\partial\phivar}\,
\varsigma\!\left(\kbt(\Phi_\xi(\phivar),\xi)\sep\Phi_\xi(\phivar),\xi\right)-
\varsigma\!\left(\kbt(\phivar,0)\sep\phivar,0\right)\right]\,,
\label{eq:sig9}
\eeq
which again solves the problem.

In summary, there are three possible ways out. The first, which we
call a type-I solution, is that of choosing a phase-space
parametrisation such that \eq{kbcond} is fulfilled (this is
essentially what has been done in ref.~\cite{Jezo:2015aia}). The
second (called type-IIa solution) entails a re-mapping of the
phase-space variables relevant to the counterevent, so that
\eq{kbcond2} is fulfilled. Finally, with a type-IIb solution
the re-mapping acts on the variables relevant to the event, 
so that \eq{kbcond3} is fulfilled. 
While approaches of type I are simpler than those of type II, 
they are not necessarily more convenient in the context of NLO or NLO+PS
computations, where the primary concern is that of finding a phase-space
parametrisation which is ideally suited to IR subtractions (and, in
the case of NLO+PS, to MC matching). The latter requirement might render 
\eq{kbcond} difficult to achieve. It also implies that it is
hard to find a solution to the problem which can be applied to a
generic IR-subtraction formalism. It is much more convenient to work 
in the context of a specific subtraction scheme, and for this reason in 
what follows we shall concentrate on FKS and on its implementation in \mga.

\subsubsection{Treatment of resonances in FKS\label{sec:genFKS}}
\begin{figure}[!ht]
\begin{center}
  \includegraphics[trim=0 0 210 600,clip,width=0.45\textwidth]{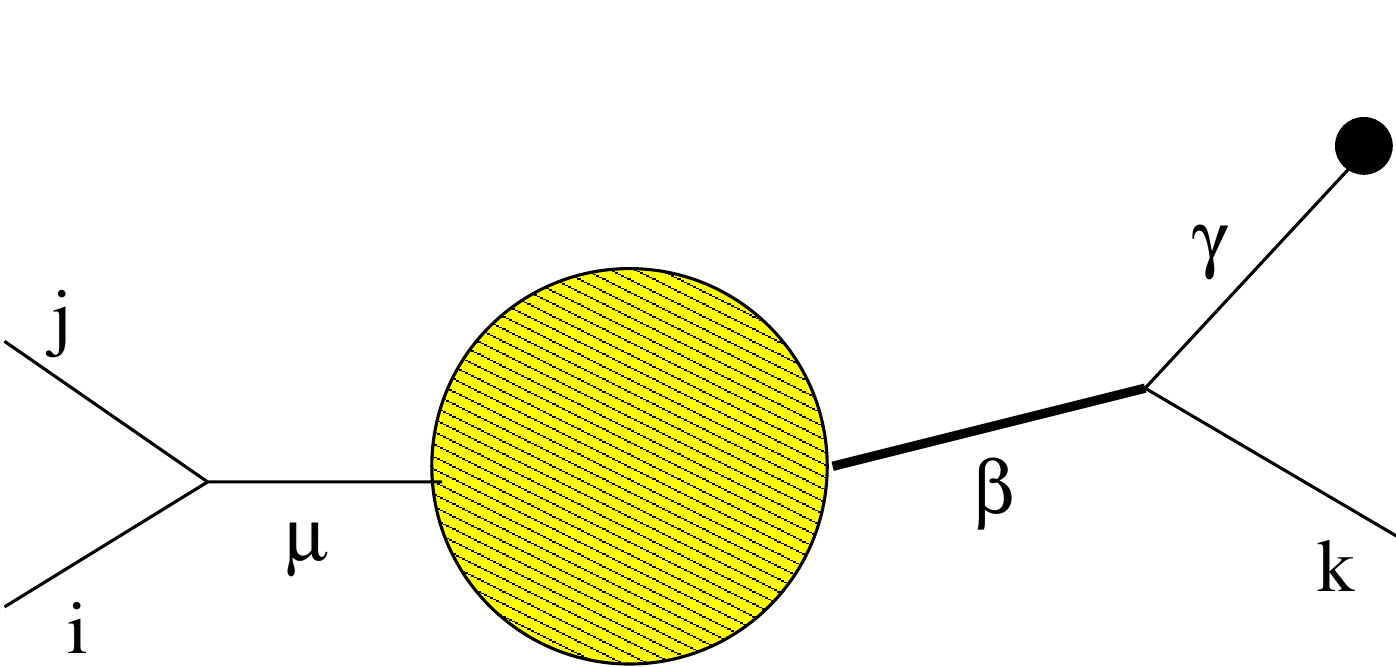}
$\phantom{aaaaa}$
  \includegraphics[trim=0 0 220 600,clip,width=0.45\textwidth]{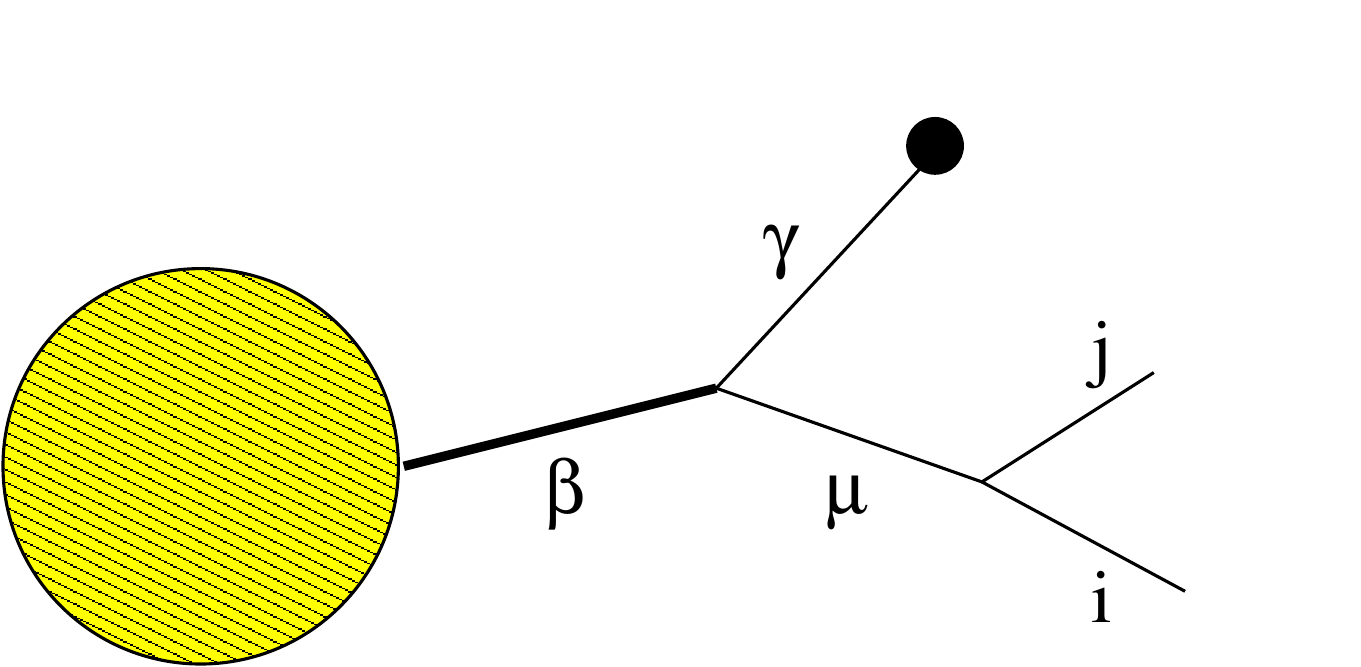}
\end{center}
\caption{\label{fig:diag1} 
Left panel: the FKS pair $(i,j)$ is not connected to the resonance $\beta$.
Right panel: the FKS pair is connected to the resonance.
}
\end{figure}

We assume that the reader is familiar with the basics of FKS subtraction;
if not, all of the relevant information can be found in the original
papers~\cite{Frixione:1995ms,Frixione:1997np} and in 
\mycite{Frederix:2009yq}. The latter work deals specifically 
with the issues relevant to automation, and hence to the
\mga implementation.
There are two possible situations, depicted in \fig{diag1}:
the case where the FKS pair -- identified in what follows by the indices
$i$ (the FKS parton) and $j$ (its sister) -- is not directly connected
to the tree\footnote{Note that this is a sensible definition, because
$\beta$ is an $s$-channel, and hence it is the root of a tree that can
be separated from the rest of the diagram by a single cut.} that stems 
from the resonance $\beta$ (left panel); and the case where the FKS pair
is part of the tree whose root is $\beta$ (right panel). 
%%%%%%%%%%%%%%%%%%%%%%%%%%%%%%%%%%%%%%%%%%%%%%%%%%%%%%%%%%%%%%%%%%%
%%%%%%%%%%%%%%%%%%%%%%%%%%%%%%%%%%%%%%%%%%%%%%%%%%%%%%%%%%%%%%%%%%%
We also have to keep in mind that, at variance with the simplified
treatment presented in \sec{gen}, in QCD there are both soft and 
collinear singularities. However, one of the key properties of the phase-space
parametrisations relevant to FKS in \mga is that for a given
real-emission resolved configuration, the reduced (i.e.~Born-like)
configurations associated with the soft and collinear limits are identical 
to each other. We observe that this is a sufficient condition for 
a type-IIb approach to work (since the re-mapping of \eq{kbcond3}
requires that the r.h.s.~of that equation be unique for a given $\phivar$).
Conversely, type-IIa solutions might be implemented in any case,
however with possibly different re-mappings associated with soft 
and collinear configurations.

The situation depicted in the left panel of \fig{diag1} can occur with 
either initial-state or final-state singularities, and the phase-space
parametrisation in \mga\ offers a type-I solution in this case. This is
because for both types of singularities all of the final-state momenta
relevant to a given event (excluding $i$ and $j$ in the case of a final-state
singularities) are related to those of the associated counterevents by means
of boosts. Since neither $i$ nor $j$ contribute to the invariant mass of
$\beta$, this implies that \eq{kbcond} is fulfilled, and therefore one can
choose $\kbt$ as an integration variable.

Let us now turn to the situation depicted in the right panel of 
\fig{diag1}, which occurs solely in the case of final-state
singularities. Relevant cases are for example that of a $Z$ branching,
with $(\beta,\mu,\gamma)=(Z,q,\bar{q})$ and $(i,j)=(g,q)$, or that of 
a top-quark branching, with $(\beta,\mu,\gamma)=(t,b,W)$ and $(i,j)=(g,b)$.
In the current version of \mga, the phase-space parametrisation adopted
is that of section~5.2 of \mycite{Frixione:2007vw}, and its generalisation
to the case of a massive FKS sister. Such a parametrisation does not obey 
\eq{kbcond}, and we have therefore considered type-II approaches.
In order to keep the present discussion at a non-technical level, we 
limit ourselves here to saying that for the sake of this work, and 
for future versions of \mga, we have implemented a type-IIb solution.
However, further details are provided in \app{reso}.

\subsection{Matching to parton showers\label{subsec:matching-to-ps}}
The matching of matrix elements to parton showers in the presence
of coloured intermediate resonances in $s$-channels presents some 
non-trivial features irrespective of the perturbative order at which 
it is carried out. In order to simplify the following discussion we
shall always refer to the process we study in this paper; however, it
should be clear that our considerations and procedure are valid in general.
At the parton level, simulations for off-shell non-resonant single-top
hadroproduction are based on processes of the type $xy\to Wbq(+z)$, 
rather than on their on-shell analogues $xy\to tq(+z)$. Therefore, 
there is no physical way (nor \emph{formal} necessity) of flagging
a specific subset of generated events as stemming from top-quark
contributions.  Nonetheless, despite the formal categorising of events as
containing or not containing an intermediate top being unphysical, the
description of higher-order contributions induced through parton showering 
might be very different in the two cases.

MC event generators typically handle the showering from a coloured resonance
and from its decay products in a factorised fashion: emissions from the
resonance are treated first and are in competition with all other sources of
radiation, and emissions off the resonance decay products are added in a
second, separate step. This choice is physically motivated by the 
NWA, which dictates a factorisation into production and decay
subprocesses, as well as a suppression of the interference of radiation in
production and in decay by $\Gamma_t/m_t \ll 1$.  In the presence of a
top-quark resonance, the showers from its decay products will usually be
forced to preserve the reconstructed top invariant mass $\mtrec$ (to be
precisely defined below). On the other hand, no such constraint is applied 
if the information on the intermediate top quark is absent. Such a disparity
may lead to very different shower evolutions even when starting from exactly
the same final-state kinematics.  For certain observables, especially
those related to the invariant mass of the $Wb$-jet system, $Wbq$ samples
for which the top quark is not written in any of the events may thus produce
results in visible disagreement with analogous on-shell $t(\to Wb)q$ samples,
even in the narrow-width limit. The reconstructed top-quark mass is itself a
prime example of this issue. This situation is disturbing, since it is
ultimately due to {\em arbitrary} choices made in MC modelling.  The decision
of whether or not to write the resonance in the event record (which is based
on the NWA that breaks down precisely in the off-shell region which one aims
to investigate), and the constraint that the resonance mass is kept constant if
written in the event record, are both choices that are non-parametric in 
nature.  These can therefore easily offset the increase in precision attained
by computing higher-order corrections.  At the same time, this implies that
systematic studies of these aspects of MC modelling can sensibly be carried
out also at the LO.

Pending thorough comparisons with data, at the theoretical level
one can assume the on-shell-top limit of the showered results to
be a sensible benchmark. Hence, the discussion above suggests that 
the explicit presence of the intermediate top in a $Wbq$ sample 
is a desirable feature under certain conditions. The most straightforward
example of the latter is that kinematic configurations for which $\mtrec$ 
is ``sufficiently close'' to the pole top mass $m_t$ should include an 
intermediate top quark at the level of hard events to be given
in input to the shower. Evidence 
for this feature comes from the fixed-order results, where
it can be shown that near top-quark resonances, the dominant contributions to
the cross section arise from Feynman diagrams involving intermediate top
quarks \cite{Papanastasiou:2013dta,Falgari:2010sf,Falgari:2011qa}.  It would
appear consistent that the leading topologies at fixed-order
should also be the dominant ones after parton-showering, and the presence of
intermediate top quarks in $Wbq$ samples would allow for this.

The strategy we have adopted in \mga\ works essentially in the same way at the
LO (where it was already the default~\cite{Maltoni:2002qb}) and at the NLO.
It is completely general, and is not restricted to the $Wbj$ process we are
considering in this paper. Moreover, it is already used, in a simplified
format, for resonances that are not charged under QCD. The method relies on
the adaptive multi-channel integration, in which integration channels are
roughly in one-to-one correspondence with Born-level Feynman
diagrams\footnote{Even in the NLO mode of \mga, the integration
  channels correspond to the underlying Born diagrams only -- see section~6.3
  of \mycite{Frederix:2009yq}, with $f=1$.}. When considering a single
integration channel, the corresponding diagram has a well-defined structure,
possibly with intermediate $s$-channel contributions.  We use this information
to decide whether any of these intermediate resonances is written in the
hard-event record. In particular, for each of such resonances:
\renewcommand{\labelenumi}{\alph{enumi}.)} 
\begin{enumerate}
\item If the diagram that defines the integration channel does not feature the
  intermediate resonance in an $s$-channel configuration, then that resonance is
  not written in the hard-event file.
\item If the diagram contains a resonance in an $s$-channel configuration,
  then we distinguish various cases depending on the FKS sector which is
  presently integrated. If the diagram and the FKS sector are such that the
  grandmother of the real-emission radiation can be identified with that
  resonance (this is the situation depicted in the right panel of \fig{diag1})
  the resonance is written on the hard-event record, and a type-IIb re-mapping
  is used, as outlined in \sec{gen} and \app{reso}, {\em unless} the MC one
  matches to does not conserve the reconstructed resonances mass when
  emissions of this kind occur.  In this same FKS sector there are
  configurations for which the re-mapping cannot be performed (see
  \app{reso}), and for these no information on the resonance appears in the
  hard event. For all of the other FKS configurations or at the LO, we use 
  a dimensionless number, $\xcut$, to determine whether the resonance will 
  be part of the hard-event record. In detail, if and only if the given 
  kinematic configuration satisfies the condition 
\beq
  \left|\mbrec-\mb\right|<\xcut \Gb
\label{eq:xcut}
\eeq
  will the resonance be part of the hard-event record. The quantity
  $\mbrec$ denotes the reconstructed mass of the resonance $\beta$,
  and is {\em defined} as the invariant mass of the four-momentum
  sum of all the decay products of the resonance, as determined by the diagram 
  used in the multi-channel integration (and the FKS sector in the case of
  real-emission-like events); $\xcut$ is a free parameter.
\end{enumerate}
\renewcommand{\labelenumi}{\arabic{enumi}.} 

\noindent
What is done here is not motivated by considerations of numerical stability,
as in the fixed-order NLO (fNLO henceforth~\cite{Alwall:2014hca}) computation
(stability is merely a by-product in the present case), but rather by
the fact that it is in keeping with the procedures adopted internally by the
MCs.  This implies that the construction of the MC counterterms relevant to
the MC@NLO matching is modified in order to take the above information into
account. In particular, the features described in items a.) and b.)  are
applied to the MC counterterms as well, in order to precisely mimic the
kinematic constraints imposed by the PS at ${\cal O}(\as^{b+1})$.  It should
thus be clear that the only new ingredient w.r.t.~the current implementation
in \mga\ is one relevant to emissions off the resonance decay products, 
which, being a phase-space re-mapping, can be trivially automated.

A couple of comments on the procedures adopted in MCs are in order here.
Firstly, for emissions that do not involve $s$-channel resonances
(left panel of \fig{diag1}) MCs can easily, and do in the cases considered
in \mga, conserve $\mbrec$. Therefore, the dependence of physical
observables on the parameter $\xcut$ is not induced by the conservation
or the lack thereof of the reconstructed resonance mass, but by the
different behaviours of the showers depending on the presence of the
resonance in the event record -- we shall see explicit examples of
this in \sec{results}. Secondly, and as far as emissions off
the resonance decay products are concerned (right panel of \fig{diag1}),
in the case of \herwig, $\mbrec$ is conserved, whilst in \pythia\ it 
is not (owing to global recoil). In the \mga\ matching to \pythia, the 
global-recoil strategy is adopted in the latter, which allows one to simplify
the MC counterterm definition in the former. In particular, the presence of
resonances becomes irrelevant in this respect, since radiation from the decay
products is generated identically to any other hard-process external partons,
which means that the same evolution kernels, phase-space boundaries and
kinematics are applied for similar evolution steps. Thus, all final-state
particles compensate for the momentum shifts necessary to give non-null
virtuality to the radiating parton.  In the case of top decays, this
ultimately implies that the invariant mass of the $Wb$ system (before an
emission) and the invariant mass of the $Wbg$ system (after an emission) are
not identical. For this reason, as already mentioned in general in item b.)
above, in the case of \pythia\ we do not apply a type-IIb phase-space
re-mapping. We have further explicitly checked that, at the LO, differences
between the global and local (which is $\mtrec$-preserving) recoil schemes are
negligible for all differential distributions studied in the context of this
paper (a sample of which will be presented in \sec{results}).

We remark that one can envisage the possibility of introducing an $\xcut$
dependence for emissions off the decay products as well (i.e.~of using
the condition of \eq{xcut} to decide whether or not to perform
a re-mapping). We have presently refrained from doing so for a couple
of reasons. Firstly, it is technically more complicated at the 
matrix-element level. Secondly, this effort would not really be justified 
in view of the fact that the Breit-Wigner function is steeply falling, 
and thus even for relatively small values of $\xcut$ one is actually quite 
close to the asymptotic case $\xcut\to\infty$. Because of this, in what
follows \herwig-based predictions have been obtained by setting
$\xcut=35$ (our default, with 35 being our arbitrary choice of a
very large value, consistently with the Breit-Wigner lineshape),
and compared to those obtained with $\xcut=0$ (in which case, we 
disallow the type-IIb re-mapping in the phase space; this option will
not be made available in the public version of \mga). On the other
hand, with \pythia\ these precautions are not necessary, and several
values of $\xcut$ will be considered. 

While suitably writing the top quark in the hard events addresses the 
on-shell-limit issue, it poses another problem in the context of the \mcnlo 
matching. Despite the top quark not being an external particle in off-shell 
$Wbq$ production, the factorised-shower structure described above allows 
MCs to radiate gluons both off the $Wb$ system \emph{and} off the top.  
The latter radiation may spoil the formal NLO accuracy of the computation. 
In fact, since gluon emission from an intermediate resonance is not 
IR-singular, in the context of the \mcnlo\ approach it is not associated 
with an MC counterterm, whence a potential double counting with the 
radiation of MC origin mentioned before.

Two solutions to this double-counting issue are possible: either a finite
extra MC counterterm is added to the \mcnlo short-distance cross section in
order to match at the NLO level the effects of the radiation off the top
quark, or this type of radiation is directly disallowed in the PS. For
simplicity, and without loss of formal accuracy, we have chosen the latter
alternative\footnote{We are grateful to Bryan Webber for providing us with a
  version of \herwig that disallows radiation off intermediate top quarks.},
but have nevertheless verified (by switching it on and off, which at least at
the LO is fully consistent) that the impact of such top-quark radiation is
negligible for all of the observables considered in this paper. The issue of
double counting is specific to the NLO, and therefore no special measures
regarding radiation from intermediate top quarks need to be taken at the LO; 
in our LO simulations, top quarks have been allowed to radiate.

\section{Results\label{sec:results}}

\subsection{Single-top hadroproduction: process definition 
and approximations\label{sec:procdef}}
The starting point of the present work is the fNLO calculation for 
the $t$-channel single-top cross section in the five-flavour scheme presented 
in \mycite{Papanastasiou:2013dta}, in which the reaction considered 
was\footnote{We emphasise that even if in \mycite{Papanastasiou:2013dta} and
in this paper we simulate $W^+$ (i.e.~top) production, the case of $W^-$ 
(i.e.~antitop) production is fully identical.}
\beq
p \; p \to W^+ \; J_b \; J_{\text{light}} + X.
\label{eq:procdef}
\eeq
As in typical single-top searches, at least two jets are present in the final
state we study: a $b$-jet, $J_b$, defined as a jet containing the outgoing
$b$-quark from the hard interaction (for this reason, we call this jet 
the \emph{primary} $b$-jet), and an additional light jet,
$J_\text{light}$, that does not necessarily contain bottom quarks.  The
assumption of a third-generation diagonal CKM matrix ($V_{tb}=1$) is made 
in order to have a self-consistent definition of the $t$-channel 
process~\cite{Papanastasiou:2013dta}\footnote{The inclusion of the 
$s$-channel contribution would require a refinement to the definition 
of the $b$-jet in order make the process well-defined. Moreover, note that
the requirement that the production be EW implies that no Born-level channel 
features a gluon in the final state, regardless of the value of $V_{tb}$.}. 
We remark that a consistent treatment of finite-width and non-resonant effects
for top-quark production at NLO can be achieved through the use of the
complex-mass scheme~\cite{Denner:1999gp,Denner:2005fg}. This is a
renormalisation scheme that introduces the top-width parameter $\Gamma_t$ as
part of a complex top-quark mass at the Lagrangian level.  Examples of recent
applications in NLO calculations with full off-shell effects are the NLO
results for top-pair~\cite{Bevilacqua:2010qb,Denner:2010jp,Denner:2012yc,
Frederix:2013gra,Cascioli:2013wga,Heinrich:2013qaa} and $t$-channel
single-top~\cite{Papanastasiou:2013dta} production.  In \mga, the generation
of the process of \eq{procdef} is obtained by issuing the following commands
(see \mycite{Alwall:2014hca} for details on the syntax):

\noindent
{\tt ~./bin/mg5\_aMC}

\noindent
~~\prompt\ {\tt ~import model loop\_sm-no\_b\_mass}

\noindent
~~\prompt\ {\tt ~set complex\_mass\_scheme True}

\noindent
~~\prompt\ {\tt ~define p = p b b\~{}; define j = p}

\noindent
~~\prompt\ {\tt ~generate p p > w+ b j \$\$ w+ w- z a QED=3 QCD=0 [QCD]}

\noindent
~~\prompt\ {\tt ~output; launch} 

\noindent
As was discussed in \mycite{Papanastasiou:2013dta}, a consistent
definition of the process of \eq{procdef} in the five-flavour scheme requires
a kinematic cut on the primary $b$-jet transverse momentum, which is not
necessary when off-shell effects are neglected, i.e.~when the top quark 
is taken to be stable. This constraint, which is relatively straightforward
to impose when working at fixed order, becomes more complicated in the
presence of parton showers, since it may become far from obvious which 
of the $B$-hadrons in the final state descends from the hard-interaction
$b$-quark. In order to directly compare our present results to previous work 
at fNLO~\cite{Papanastasiou:2013dta}, and to ensure that the conclusions 
presented below at the hadron level are in as close analogy as possible 
with those of ref.~\cite{Papanastasiou:2013dta}, we have chosen to exploit 
MC truth\footnote{In both the \herwig and \pythia showers, the 
mother (\texttt{JMOHEP}) and daughter (\texttt{JDAHEP}) arrays in the event 
record are used to perform this identification. In the \herwig analysis we 
also make use of the information on the space-time vertices (\texttt{VHEP}) 
where particles are produced.}. This enables the tagging of the primary 
$B$-hadron and thus that of the primary $b$-jet, the latter being identified 
as the jet containing the primary $B$-hadron. 
In addition to the $\pt$ cut on the primary $b$-jet, we impose other 
cuts at the analysis level, summarised in \tab{cuts}, which we adopt 
throughout our simulations. They are identical to those of
\mycite{Papanastasiou:2013dta}, and thus allow for a direct comparison
with that paper. We limit ourselves here to reminding the reader that this
setup has been chosen so as to avoid artificial enhancements of 
non-resonant contributions, so that a meaningful comparison with the 
on-shell single-top simulations can be made.  
%%%%%%%%%%%%%%%%%%%%%%%%%%%%%%%%%%%%%%%%%%%%%%%%%%%%%%%%%%%%%%%%%%%
\begin{table}[h]
\centering
\begin{tabular}{l l}
\hline \\[-5pt]
$\pt(J_b) > 25 \text{ GeV}$ & $\pt(J_{\text{light}}) > 25 \text{ GeV} $ \\[5pt]
$\lvert \eta(J_b) \rvert < 4.5$ & $\lvert \eta(J_{\text{light}}) \rvert < 4.5$ \\[5pt]
\multicolumn{2}{c}{$140~\mbox{GeV} < M(W^+,J_b) < 200~\mbox{GeV}$} \\[5pt]
\multicolumn{2}{c}{$k_t$ jet algorithm \cite{Catani:1993hr,Cacciari:2011ma}, with $R_{\text{jet}}=0.5$} \\[5pt]
\hline
\end{tabular}
\caption{Setup for process definition and analysis.}
\label{tab:cuts}
\end{table}
%%%%%%%%%%%%%%%%%%%%%%%%%%%%%%%%%%%%%%%%%%%%%%%%%%%%%%%%%%%%%%%%%%%

Given that observables sensitive to the leptonic decays of the $W$ boson are
also of interest in single-top analyses (e.g.~the invariant mass of the
lepton+$b$-jet system, $M(l^+,J_b)$, or the transverse mass of the
reconstructed top quark, $M_T(l^+,\nu_l,J_b)$), we consider a leptonically
decaying $W^+$. For events passed to the PS, these decays are carried out by
the showers themselves, whereas at fixed order we simply decay the $W^+$
isotropically in its rest frame (which is also what the MCs do). We note that
in this way production spin correlations for the leptons are not included. 
While these may be phenomenologically important, they are not relevant for 
the assessment of the off-shell and non-resonant effects we are presently 
interested in, and therefore do not warrant the more involved process 
definition that would be necessary when generating directly leptonic 
matrix elements. 

In view of the process definition and of the cuts in \tab{cuts},
one expects its on-shell analogue to constitute a reasonable approximation.
Indeed, as it has been shown both in single top~\cite{Falgari:2010sf,
Falgari:2011qa,Papanastasiou:2013dta} and in $t\bar{t}$~\cite{Denner:2010jp,
AlcarazMaestre:2012vp,Falgari:2013gwa} production, the NWA does an excellent 
job in approximating the fully-off-shell results for many distributions.
However, since it does fail to capture the dominant effects in regions of
phase space that are sensitive to the top off-shellness and to non-resonant
contributions, misuse of the NWA may therefore introduce errors that vastly
exceed the na\"ive estimate of $\sim \mathcal{O}(\Gamma_t/m_t)$, which
ultimately may have a bearing on experimental procedures such as top
tagging and top-mass extractions.

The presence of potentially large effects of this kind will be studied
in the following through a systematic comparison of the $W^+bj$ results
with their on-shell counterparts, which we generically denote by $tj$.
In $tj$ production the top is a stable external particle, hence its 
radiation in the shower is consistently matched  at the level of the \mcnlo 
short-distance cross sections. Thus, in this case, showering from the 
top \emph{must} be allowed in order to avoid double counting: this is
the usual procedure~\cite{Frixione:2005vw,Frederix:2012dh}. 
The inclusion of NLO corrections to $tj$ production, while highly 
desirable, is largely incomplete from the phenomenology viewpoint,
since it does not improve the description of the top decays (which is 
left to the MC), and does not include any non-resonant contributions. 
This situation is addressed in part by the use of the procedure
of \mycite{Frixione:2007zp}, automated in \mga\ in the module
\ms~\cite{Artoisenet:2012st}, which allows one to include both production 
and decay spin correlations, and to give a rough description of off-shell
effects through a simple Breit-Wigner smearing. For this reason, in
the following we shall always use $tj$ predictions in
conjunction with \ms. Although an improvement w.r.t.~the ``bare''
$tj$ results, these still do not contain NLO corrections to top 
decays, and fully ignore the non-resonant contributions to the 
$W^+bj$ final state. 

In summary, by systematically comparing fNLO, $tj$+\ms\ NLO+PS, 
and $W^+bj$ NLO+PS results, as well as their LO counterparts, we 
shall be able to assess the impact of a variety of mechanisms, since
the above simulations are characterised by an increasing degree of 
complexity, owing to the inclusion of parton showers, of NLO corrections
to decays, and of off-shell and non-resonant contributions.

\subsection{Differential distributions}
In this section we present our predictions for several observables, obtained
with the different computational schemes discussed in \sec{procdef}. All of
these have been derived by setting the input parameters that enter the hard
matrix elements as shown in \tab{parameters}. The width value labelled by
``NLO'' in \tab{parameters} is adopted in the context of the $W^+bj$ NLO+PS
and fNLO calculations, whereas that labelled by ``LO'' is used in all the
other cases. Theory uncertainties are estimated by varying the 
renormalisation and factorisation scales independently in
the range $[\mu_0/2,2\mu_0]$. These variations are performed automatically 
by \mga\ in the course of a single run, thanks to the re-weighting
technique introduced in \mycite{Frederix:2011ss}\footnote{This implies 
that the scale dependence of the top width is neglected.}. We point out 
that, for certain observables, the LO scale dependence may be pathologically
small, since the Born cross section does not contain any $\as$ factor.
We have refrained from reporting the uncertainties associated with PDF errors,
chiefly in view of the fact that they affect equally the $W^+bj$ and $tj$
production processes. We have run the two MCs by adopting the respective
default parameters, except for the PDFs, which have been taken equal to
those used in the short-distance computations. The simulation of the
underlying events is turned off, and in order to simplify the analysis
$B$ hadrons are imposed to be stable.
%%%%%%%%%%%%%%%%%%%%%%%%%%%%%%%%%%%%%%%%%%%%%%%%%%%%%%%%%%%%%%%%%%%
\begin{table}[h]
\centering
\begin{tabular}{l l}
\hline \\[-7pt]
$m_Z = 91.1876 \text{ GeV}$ & $m_W = 80.3980 \text{ GeV}$ \\[5pt]
$\Gamma_Z = 2.4952 \text{ GeV}$ & $\Gamma_W = 0 \text{ GeV}$ \\[5pt]
$G_F=1.6639 \times 10^{-5} \text{ GeV}^{-2} $ & $\alpha^{-1}_{e}=132.3384$ \\[5pt]
$m_t = 173.2 \text{ GeV}$ & $m_b = 0 \text{ GeV}$ \\[5pt]
$\Gamma^{\text{LO}}_t = 1.5017 \text{ GeV}$ & $\Gamma^{\text{NLO}}_t(\mu=m_t/2)=1.3569 \text{ GeV}$ \\[5pt] 
Central scale: $\mu_0 = m_t/2$ & PDFs: \texttt{MSTW2008NLO} \cite{Martin:2009iq} \\[5pt]
\hline
\end{tabular}
\caption{Input parameters for hard matrix elements.}
\label{tab:parameters}
\end{table}
%%%%%%%%%%%%%%%%%%%%%%%%%%%%%%%%%%%%%%%%%%%%%%%%%%%%%%%%%%%%%%%%%%%

In the following, to each observable we associate a figure that contains two
main panels, one for \herwig and one for \pythia, each accompanied by three
insets.  The main panels display four curves: $W^+bj$ results at NLO+PS and
LO+PS with $\xcut=35$ (solid and dashed blue, respectively), at fNLO (dashed
green with full diamonds), and $tj$+MS results at NLO+PS (solid red with full
circles). \ms decays are characterised by a user-defined parameter, \bwcut,
that sets the allowed range (i.e.~the distance, in width units, from the
resonance pole mass) for the invariant mass of the system composed of the
resonance decay products.  We choose $\bwcut=35$ as our default, and indicate
this explicitly by appending the value of \bwcut~to the label `MS' in the
plots (e.g.~`MS35' indicates \ms results with $\bwcut=35$, and so on). We
emphasise that the parameters $\xcut$ and \bwcut~are technically different,
even though they are both associated with a distance from the resonance pole
mass.  \ms simulations feature a top quark in \emph{all} of their Les-Houches
events and therefore are independent of $\xcut$ (more precisely, they are
characterised by $\xcut=\infty$) but depend on \bwcut.  Conversely,
simulations of the full $W^+bj$ process do not require MS decays, thus these
events are strictly \bwcut-independent, but do carry a dependence on $\xcut$.
Still, $\xcut$ and $\bwcut$ have a similar meaning from a physics viewpoint,
because they parametrise, in different contexts, effects related to top-quark
off-shellness. This is the reason why we have chosen their default values to
be identical. Similarly, $\bwcut=0.1$ is the analogue\footnote{In \ms\ one
  cannot set $\bwcut$ strictly equal to zero, and thus we have used $0.1$
  instead. Given the values of the top mass and width, this difference is
  fully irrelevant.}  of $\xcut=0$.

The first (upper) inset in each figure contains ratios of the various
perturbative approximations to the full $W^+bj$ process: LO+PS/NLO+PS (solid
blue), fLO/fNLO (dashed green), \herwig/\pythia at NLO+PS and at LO+PS (solid
and dashed brown on the right, respectively), in addition to scale-variation
bands (\herwig only, LO in yellow, NLO in grey on the left).  The second
(middle) inset contains the ratio, with respect to NLO+PS $W^+bj$ with
$\xcut=35$, of NLO+PS $tj$+MS with $\bwcut=35$ and with $\bwcut=0.1$ (solid
red and dashed magenta, respectively) and of fNLO (dashed green).  Finally,
the third (lower) inset displays the ratio, with respect to NLO+PS $W^+bj$
with $\xcut=35$, of NLO+PS $W^+bj$ with $\xcut=$ 0, 1, and 5 (solid cyan,
dashed red, and solid green, respectively, with the latter two values
adopted only for \pythia, as explained in \subsec{matching-to-ps}).

In order not to further complicate the discussion of the different effects
that play a role in the results presented below, for each figure our analysis
is structured as follows. After a brief overview of the observable examined
and the important features of the fixed-order results (including the effects
of NLO corrections, sensitivity to off-shellness, and non-resonant 
contributions), we address in turn:
\begin{enumerate}%\itemsep3pt \parskip0pt \parsep0pt
\item the effect of NLO corrections on the matched results for the full
  $W^+bj$ process (first inset, solid blue curve);
\item the effect of parton showering with respect to the fixed-order results
  for the full $W^+bj$ process (first and second insets);
\item the differences between results for the full $W^+bj$ process, showered
  with \herwig and \pythia (first inset, dashed and solid brown curves
  on the right-hand panels);
\item the quality of the stable-top+\ms approximations to the full result
  (second inset, solid red and dashed magenta curves);
\item the sensitivity of the results to the arbitrary $\xcut$ parameter (third
  inset).
\end{enumerate}

\subsubsection{Transverse momentum of reconstructed top quark, $\pt(W^+,J_b)$}
\begin{figure}[h]
\hspace{-0.6cm}
\includegraphics[trim=0.0cm 0.1cm 0.0cm 0.1cm,clip,width=1.10\textwidth]{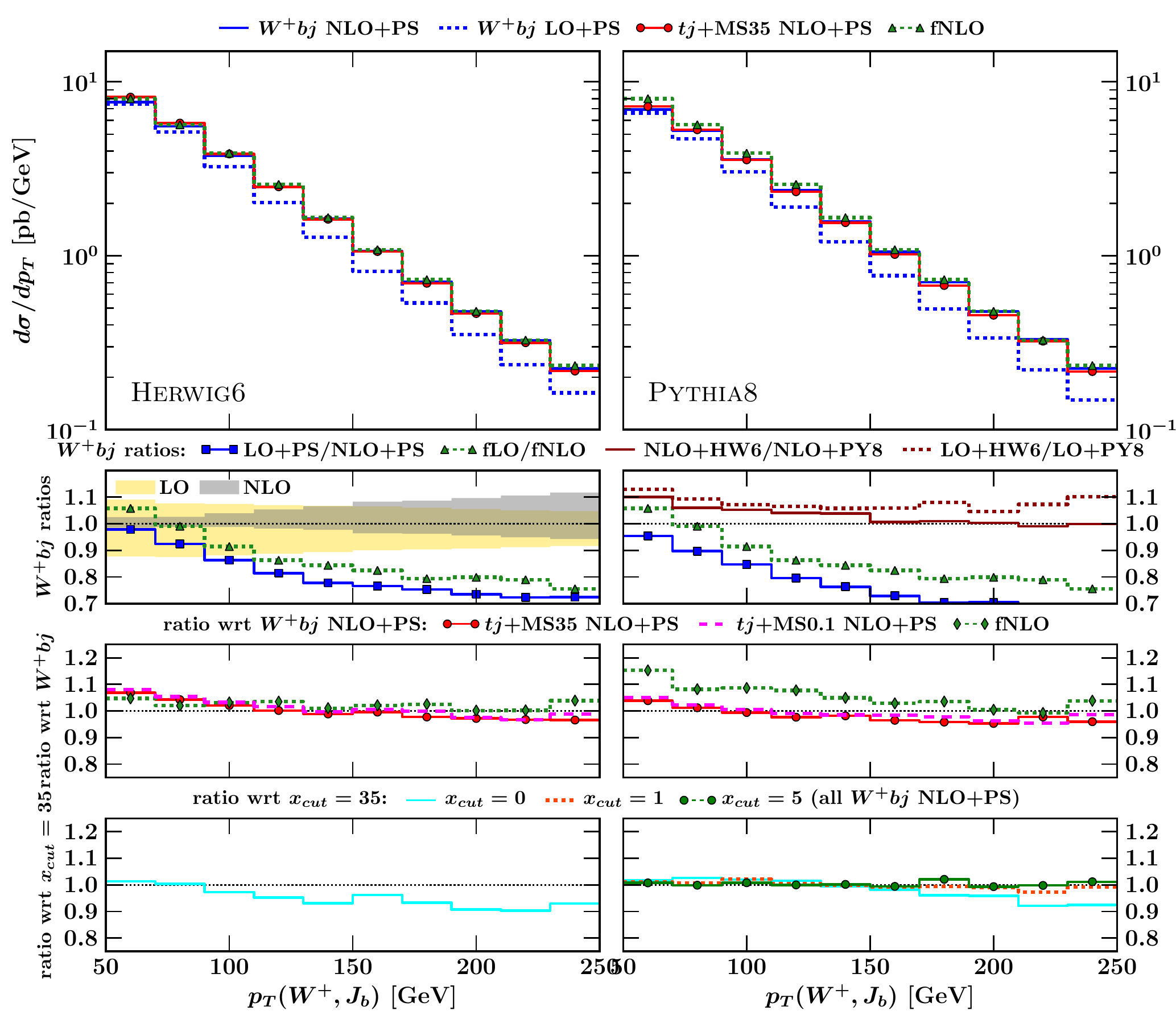} 
\caption{Transverse momentum of reconstructed top, $\pt(W^+,J_b)$.}
\label{fig:pTWb}
\end{figure}

\noindent
The first observable we examine is $\pt(W^+,J_b)$, the transverse momentum of 
the reconstructed top quark (defined as the system composed of the $W^+$ 
and the primary $b$-jet), shown in \fig{pTWb}.  This observable is
inclusive in the invariant mass of the reconstructed top and is thus barely
sensitive to off-shell effects \cite{Papanastasiou:2013dta}.
Resonant/non-resonant interferences and pure non-resonant effects also do not
play a major role.  By comparing the fNLO results to
MCFM~\cite{Campbell:2004ch} ($t$-channel single top in the NWA, with NLO
corrections in both production and decay; not shown here) it can be deduced
that the trend of the spectrum becoming harder at high $\pt(W^+,J_b)$ at fNLO
w.r.t.~fLO is a direct consequence of the corrections to production, but is
however also enhanced by those to the decay.
%%%%%%%%%%%%%%%%%%%%%%%%%%%%%%%%%%%%%%%%%%%%%%%%%%%%%%%%%%%%%%%%%%%
%%%%%%%%%%%%%%%%%%%%%%%%%%%%%%%%%%%%%%%%%%%%%%%%%%%%%%%%%%%%%%%%%%%
\begin{enumerate}%\itemsep3pt \parskip0pt \parsep0pt
\item The inclusion of NLO corrections to the matched simulation mirrors the
  hardening of the spectrum observed when including NLO corrections at a fixed
  order.
\item The dashed green curve in the second inset indicates that PS effects are
  on the whole not large (below 10\% in all bins except the first bin for
  \pythia), and do not significantly alter the shape of the fixed-order
  results.
\item The agreement between \herwig and \pythia improves at NLO (the solid
  brown curve on the first right-hand inset is systematically closer to unity 
  than the dashed brown curve). However, it should be pointed out that the
  agreement is already good at LO (especially shape-wise).
\item The second inset reveals that there is a general trend of the $tj+$MS
  spectra becoming softer compared to those of the NLO+PS $W^+bj$ results.
  This effect follows from the same reasoning as the softer behaviour of the
  fLO or the LO+PS spectra w.r.t.~their NLO counterparts, but is of much
  smaller size since the \ms results do include radiative corrections to the
  production subprocess, while they lack those to the top decay.  Overall, the
  results for NLO+PS $tj$+MS are in agreement with the full NLO+PS $W^+bj$
  ones to better than 10\% for both showers, indicating that not only
  corrections to the decay but also non-resonant contributions are small in
  this case.  Moreover, the invariance of the \ms results under variation of
  the \bwcut~parameter confirms that this distribution is relatively
  insensitive to off-shell effects.
\item The third inset reveals that $\pt(W^+,J_b)$ is largely stable against
  the choice of $\xcut$, with only $\xcut=0$ displaying any visible
  effects. The latter are however smaller than (\pythia) or comparable to
  (\herwig) the NLO scale uncertainty illustrated by the band in the upper 
  left inset.
\end{enumerate}
\noindent

\subsubsection{Transverse momentum of primary $b$-jet, $\pt(J_b)$}
%%%%%%%%%%%%%%%%%%%%%%%%%%%%%%%%%%%%%%%%%%%%%%%%%%%%%%%%%%%%%%%%%%%
\begin{figure}[h]
\hspace{-0.6cm}
\includegraphics[trim=0.0cm 0.1cm 0.0cm 0.1cm,clip,width=1.10\textwidth]{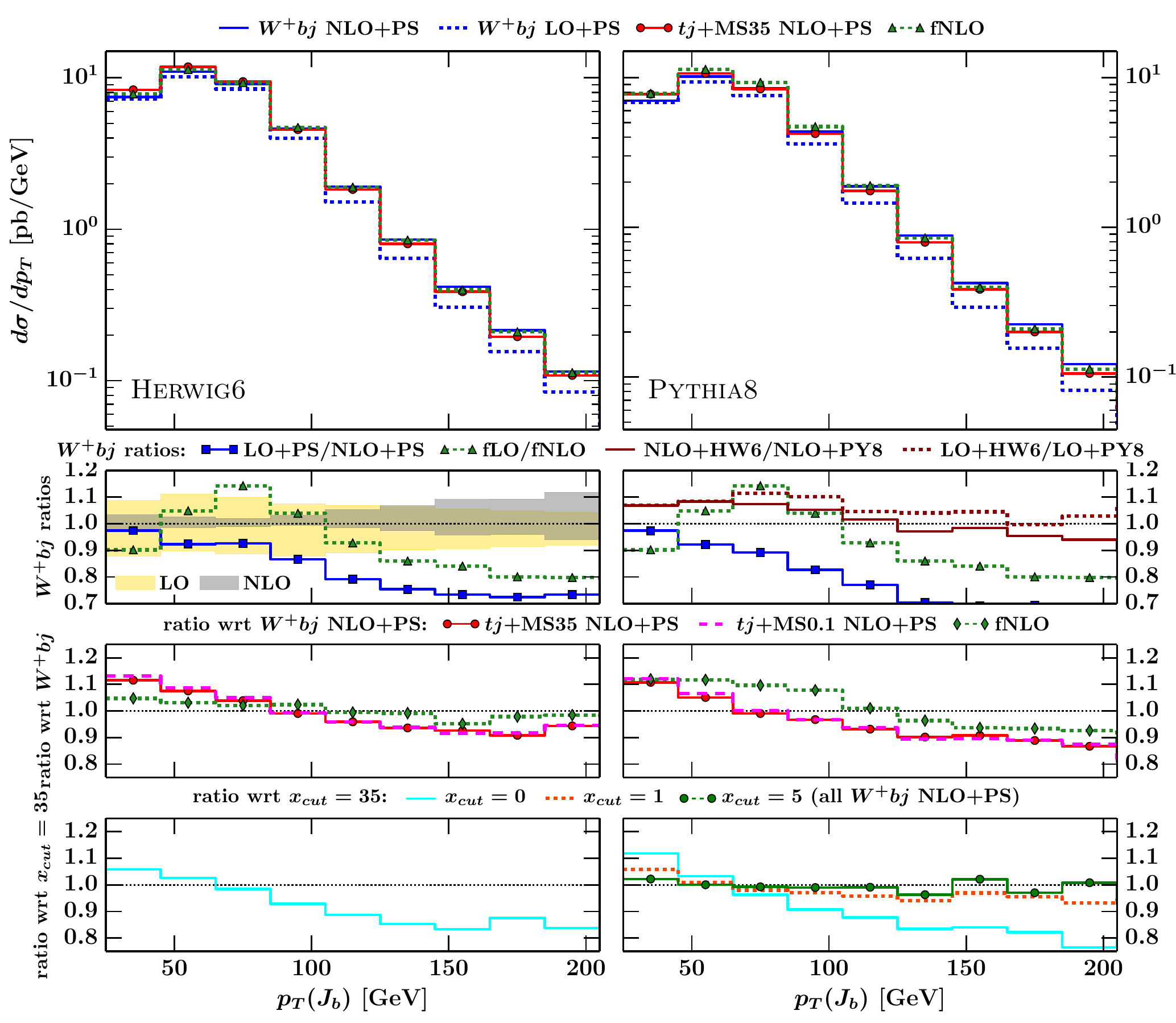} 
\caption{Transverse momentum of primary $b$-jet, $\pt(J_b)$.}
\label{fig:pTJb}
\end{figure}
%%%%%%%%%%%%%%%%%%%%%%%%%%%%%%%%%%%%%%%%%%%%%%%%%%%%%%%%%%%%%%%%%%%
\noindent
Figure~\ref{fig:pTJb} shows the transverse momentum of the primary $b$-jet,
$\pt(J_b)$. This observable is less inclusive than $\pt(W^+,J_b)$ over the top
decay products, and therefore NLO corrections to the decay are expected to
play a more important role.  This is indeed the case since the non-trivial
shape at low $\pt$ of the differential $K$-factor at fixed order is driven by
the NLO corrections to the top decay (this has again been cross-checked with
MCFM). These corrections also result in a harder $\pt(J_b)$-tail at NLO
w.r.t.~LO.  The feature at small $\pt$ in the fLO/fNLO ratio can be attributed
to the kinematical fact that real radiation off the $b$-quark carries energy
away from the $b$-jet, thus softening the NLO spectrum; such a leakage 
occurs less often when moving towards large $\pt$'s, where 
the jets tend to be more collimated.
\begin{enumerate}\itemsep3pt \parskip0pt \parsep0pt
\item The differential $K$-factors for the showered results at large values of
  $\pt(J_b)$ display the same features as the fixed-order results (NLO+PS
  distributions are harder than LO+PS ones). However, the kinematic suppression
  in fLO/fNLO at low $\pt(J_b)$, driven by the fLO shape, does not carry over
  to the showered case. This is due to the fact that the shower, already at the
  LO, accounts for multiple emissions from the final-state $b$ quark, hence the
  radiation leakage outside the $b$-jet, induced by real corrections in
  the fixed-order case, has a much milder impact at the showered level.
\item The dashed green curve in the second inset indicates that the shower
  effects for \herwig at NLO are small, i.e.~that NLO+PS is very close to
  fNLO, with a distribution only marginally harder ($\pm5\%$ at low and high
  $\pt(J_b)$, respectively).  With effects of around $\pm$10\%, \pythia
  departs more from the fNLO result.
\item From the brown dashed and solid curves in the upper right inset, we
  conclude that there are only mild shape differences between the \herwig and
  \pythia predictions; \pythia\ tends to be slightly harder than \herwig.
  The ratio of the two MC predictions displays a more regular
  behaviour (i.e.~in a larger $\pt$ range) at the NLO than at the LO.  This is
  most likely due to the fact that the impact of the matrix-element
  normalisation constraint is more important in the former than in the latter
  case. Such a pattern is similar to that observed in the case of
  $\pt(W^+,J_b)$ but is of slightly bigger size here, which is consistent with
  the fact that the present observable is more dependent on MC modelling than
  the transverse momentum of the pseudo top.
\item The trend of the NLO+PS $tj$+MS curves (solid red, second inset) closely
  follows that of the LO+PS $W^+bj$ predictions, namely they are softer
  than the NLO+PS $W^+bj$ benchmarks.  The effect is similar to that observed
  in $\pt(W^+,J_b)$ but is somewhat more pronounced here.  Given that at fNLO
  in the NWA it is the corrections to the decay that induce the dominant
  features of the fLO/fNLO ratio, and that it is precisely these corrections
  that are missing in the \ms results, this is a strong indication that
  corrections to the decay subprocess are important for this observable.
  The independence of the \ms result on the \bwcut~parameter indicates that
  off-shell effects are essentially irrelevant for this observable -- a 
  feature that can also be seen at fixed order.
\item As for the case of $\pt(W^+,J_b)$, the present observable is relatively
  insensitive to the value of $\xcut$.  We only observe a marked effect for
  $\xcut = 0$, with differences to the $\xcut=35$ result of up to 20\% in the
  hard tails, for both showers.
\end{enumerate}

\subsubsection{Invariant mass of reconstructed top quark, $M(W^+,J_b)$ 
\label{subsec:mwb}}
%%%%%%%%%%%%%%%%%%%%%%%%%%%%%%%%%%%%%%%%%%%%%%%%%%%%%%%%%%%%%%%%%%%
\begin{figure}[h]
\hspace{-0.6cm}
\includegraphics[trim=0.0cm 0.1cm 0.0cm 0.1cm,clip,width=1.10\textwidth]{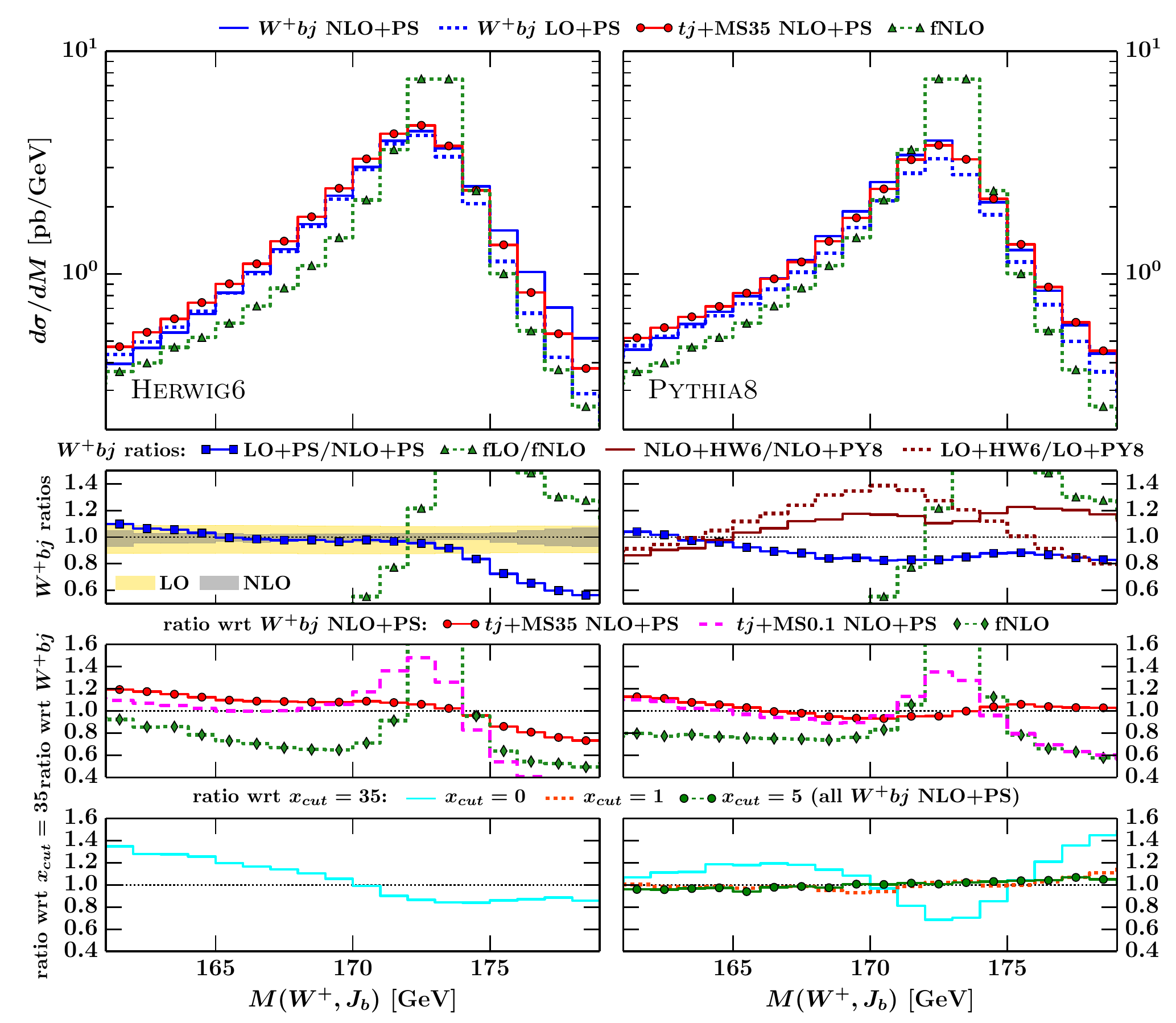} 
\caption{Invariant mass of reconstructed top, $M(W^+,J_b)$.}
\label{fig:MWb}
\end{figure}
%%%%%%%%%%%%%%%%%%%%%%%%%%%%%%%%%%%%%%%%%%%%%%%%%%%%%%%%%%%%%%%%%%%

\noindent
The reconstructed top quark mass, $M(W^+,J_b)$, displayed in \fig{MWb}, is an
important observable used to tag top quarks and to help separate the single-top
signal from its backgrounds. It may also be used in various ways to extract 
the top mass from data.  At fixed order, the real-radiation 
corrections to both production and decay are important and 
are the dominant contributions to the shape of the fLO/fNLO ratio.
The region above the peak is sensitive to radiation from the production
subprocess, whereas the region below the peak is sensitive to radiation from
the top decay products.  Additionally, treating the top quark as off-shell is
vital to sensibly describe this distribution at fixed order, with the
predictions using the NWA failing to capture most of its features;
see ref.~\cite{Papanastasiou:2013dta} for more details.
\begin{enumerate}\itemsep3pt \parsep0pt
\item The effect of the NLO corrections on the LO+PS curve is to skew the
  distribution towards the right.  The agreement in shape between LO+PS and
  NLO+PS in the low-mass region, i.e.~the region sensitive to radiation from
  the final-state $b$-quark, is satisfactory for both showers (and slightly
  better for \herwig).  This is an indication that for this observable, and in
  this phase-space region, hard radiation originating from the top-quark decay
  products is well approximated by the parton showers.  The harder spectra at
  NLO+PS, particularly visible from the large-mass slope of the \herwig
  result, stem from hard radiation in the production subprocess being
  clustered into $J_b$ by the jet algorithm.  The \pythia high-mass tail does
  not show as strong a trend as \herwig, likely pointing to more (or harder)
  production radiation in the LO results compared to the \herwig shower. The
  fact that this behaviour is mostly driven by the LO predictions can be
  inferred from the two brown histograms in the upper right-hand inset -- see
  item 3 below.
\item The effect of parton showering with respect to fixed-order results is
  very significant over the full range considered, and exceeds 50\% in the
  bins near the peak. Radiation by both showers smears and flattens the
  sharply-peaked fixed-order distribution.  This smearing results from the
  combined effect of ISR and FSR enhancing the high-mass tail when clustered 
  into the $b$-jet, and of $b$-quark FSR enhancing the low-mass tail
  when leaking out of the $b$-jet.
\item The dashed and solid brown curves in the first right-hand inset indicate
  that, both at LO and NLO, the \pythia distributions are flatter overall than
  the corresponding \herwig ones, with effects as large as 20\% and 40\%
  at NLO and LO, respectively.  Despite the remaining visible
  differences, there is a substantial improvement in the consistency of the
  two showers at NLO, compatible with the increased formal accuracy of the
  simulation.  Differences between the two showers are to some extent expected
  -- the smearing cannot be attributed to one single factor, but rather is
  a combination of various sources that vary between the showers: different
  $\as(m_Z)$ or $\Lambda_{\textnormal{\tiny QCD}}$ choices, different
  showering models (interleaved ISR/FSR in \pythia and sequential ISR/FSR in
  \herwig) and different hadronisation models.
\item The results for $tj$+MS are, for large values of \bwcut, in good
  agreement (20\% or better in the case $\bwcut=35$) with the NLO+PS $W^+bj$
  distributions. This is particularly true for \pythia, especially above the
  peak, while in the case of \herwig the $tj$+MS result displays a 
  softer behaviour over the full mass range considered. As for 
  the observables considered previously, for both showers the $tj$+MS/$W^+bj$ 
  ratio has a similar pattern, though milder (i.e.~it is closer to one), as
  that of the $W^+bj$ LO+PS/NLO+PS ratio. The large discrepancies between
  the $\bwcut=35$ and $\bwcut=0.1$ results, which are significant close to the
  peak (and, to a lesser extent, above it), indicate the importance of
  including off-shell effects for a good description of this observable in
  that region.  On the other hand, all $\bwcut$ choices appear to be in
  good mutual agreement below the peak. This is likely due to off-shell 
  effects being subdominant in this region w.r.t.~corrections to the top 
  decay products.
\item $M(W^+,J_b)$ exhibits a sensitivity to $\xcut$ qualitatively similar to
  that seen for $\pt(J_b)$ and $\pt(W^+,J_b)$, namely there is a very small
  dependence for \mbox{$\xcut\ge 1$}, while the extreme choice $\xcut=0$ gives
  some visible shape distortions\footnote{This statement depends on the
    pseudo-top mass range considered. If plotted in a range wider than that of
    \fig{MWb}, the $\xcut=1$ and $\xcut=5$ results (the former to a much
    larger extent than the latter, as expected) would also exhibit
    increasingly large differences w.r.t.~the $\xcut=35$ one.}.  For both MCs
  the $\xcut=0$ results are flatter, more markedly so with \pythia, which has
  also a mild tendency to skew the distribution rightwards, while in the case
  of \herwig\ the skewing is rather leftwards. This shows that, when the MCs
  have no information about the intermediate resonance, large model-dependent
  effects can be introduced in the resonance structure.  In the region of
  $M(W^+,J_b)$ above the peak, \pythia appears to be significantly more
  sensitive than \herwig to the choice of $\xcut = 0$. This is related to the
  different construction principle underlying the showering models. While
  \herwig first generates initial-state radiation, and follows up by
  generating final-state emissions, \pythia constructs both showers in a
  combined, interleaved sequence. Thus, initial- and final-state radiation are
  in direct competition in \pythia. This competition is drastically different
  for $\xcut=0$ and $\xcut=35$.  In general, the PS can emit from the
  initial-state partons, the light final-state partons, and from the top-quark
  decay products.  In the $\xcut=0$ case, all of these possibilities compete
  with each other for phase space.  In the $\xcut=35$ case, the evolution is
  split into showering the production process, and showering the decay
  products. Thus, while all partons associated with the production process
  compete with each other, the radiation off the decay products is not
  encumbered by any competition\footnote{Note that the absence of competition
    does not mean the absence of constraints, since overall phase-space
    boundaries and momentum conservation have to be respected. For example, if
    no radiation off other legs were present, then the constraints for
    radiation off the $b$ quark would be independent of $\xcut$, in the global
    recoil scheme.}. These different mechanisms -- due to the different
  evolution prescriptions in the $\xcut=0$ and finite $\xcut$ cases -- lead to
  the $\xcut$ dependence seen in the \pythia\ results.

  It is beyond the scope of this work to decide which showering
  model (interleaved versus independent) is preferable. We have instead chosen
  to document the features related to this choice for the process under
  consideration, and regard the $\xcut$ dependence as a way of parametrising
  this modelling uncertainty. The fact remains, that by never writing the
  top on the hard-event record, one becomes more sensitive to the different
  underlying shower mechanisms.
\end{enumerate}

\subsubsection{Mass of primary $b$-jet, $M(J_b)$\label{subsec:mjb}}
%%%%%%%%%%%%%%%%%%%%%%%%%%%%%%%%%%%%%%%%%%%%%%%%%%%%%%%%%%%%%%%%%%%
\begin{figure}[h]
\hspace{-0.6cm}
\includegraphics[trim=0.0cm 0.1cm 0.0cm 0.1cm,clip,width=1.10\textwidth]{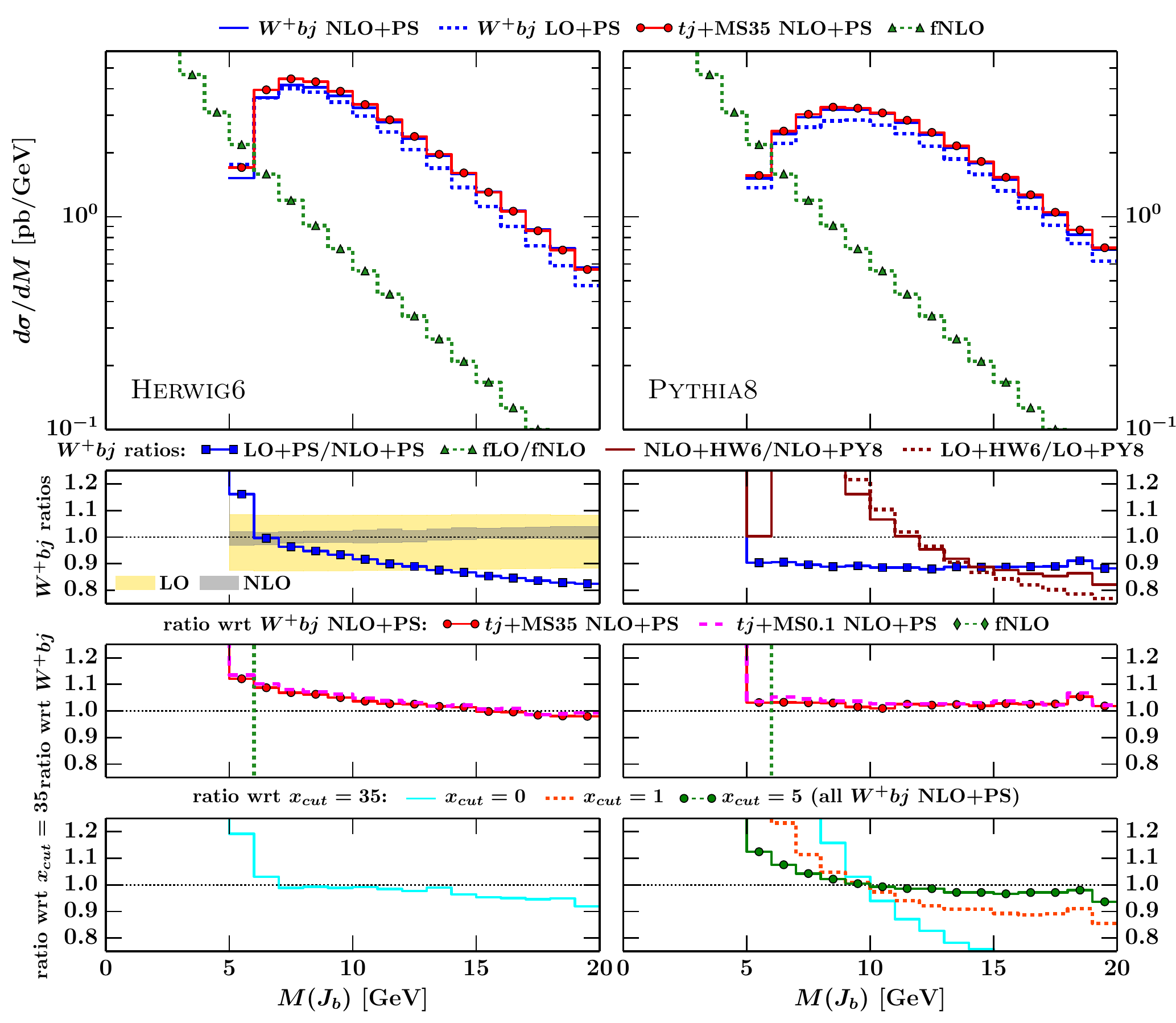} 
\caption{Mass of primary $b$-jet, $M(J_b)$.}
\label{fig:MJb}
\end{figure}
%%%%%%%%%%%%%%%%%%%%%%%%%%%%%%%%%%%%%%%%%%%%%%%%%%%%%%%%%%%%%%%%%%%

\noindent
Next, we examine the invariant mass of the primary $b$-jet, $M(J_b)$,
displayed in \fig{MJb}.  Due to the analysis setup we have adopted
(specifically the requirement of both a $b$-jet and a light jet in the final
state), at fLO $M(J_b)$ only receives contributions in the first bin,
$M(J_b)=m_b=0$.  At fNLO, the region $M(J_b) > 0$ is filled by events where
real radiation is clustered together with the $b$-quark to form the
$b$-jet. Since only real radiation contributes non-trivially in this region,
the fNLO prediction diverges as $M(J_b)\to 0$.  Shower effects are dramatic:
the threshold is shifted from zero to the mass of the lowest-lying $B$ hadron,
and the low-mass divergence present at fixed order is offset by the usual
Sudakov damping.  Consequently, discrepancies in this case are expected to be
significant, both in the comparison between NLO+PS and fNLO results, and
between different MCs.  We point out that the characteristics of the present
observable outlined above render it analogous to any quantity which has only
kinematically-trivial contributions at fLO, meaning that it displays maximal
sensitivity to real radiation and to the shower. In these cases, a by-product
of matching to showers is also that of featuring an NLO-type scale uncertainty
in the region which receives solely non-hard real-emission
contributions. This analogy only holds to a certain extent here since, for the
$b$-jet mass, hard (as opposed to soft) real radiation cannot be associated
with certainty to a specific region of the phase space.  For example, a hard
emission from the initial state could be nearly collinear to the final-state
$b$-quark, yielding a small $b$-jet mass. However, one expects the impact of
these configurations to be subdominant.

Jet masses will, in general, aside from an increased sensitivity to 
perturbative effects (including
NLO-matching systematics), also show a relatively strong dependence on
non-perturbative and soft-physics modelling; this is particularly true at
small masses. Therefore, any conclusion based on varying only ``perturbative
parameters'' is potentially incomplete.
\begin{enumerate}\itemsep3pt \parskip0pt \parsep0pt
\item Including NLO corrections when matching to parton showers leads to a
  harder $M(J_b)$ distribution in the case of \herwig, while induces only a
  constant shift for \pythia. In both cases the effects are relatively mild
  (up to 20\% for \herwig, and 10\% for \pythia), which is remarkable
  if compared with the situation at fixed order.
\item The effects of parton showering for $M(J_b)$ lead to completely
  different distributions with respect to fixed-order results; the latter are
  indeed not particularly sensible for an observable of this type.  As
  discussed before, this stems from two main reasons. Firstly, at fixed order
  the bins for $M(J_b)>0$ only receive contributions from real
  corrections. Secondly, $b$-jets are reconstructed at the hadron level, and
  hence their mass threshold is close to the physical mass of $B$ hadrons. We
  remark, however, that even if $b$-jets were reconstructed at the parton
  level in (N)LO+PS simulations, one would obtain a very similar threshold,
  owing to the fact that the MCs need to turn $b$ quarks into massive objects
  with $m^\text{MC}_b\sim5$~GeV, in order to give a realistic description of
  $b$-physics phenomena.
\item The comparison between \pythia and \herwig reveals sizable differences,
  compatible with the fact that this observable receives large contributions
  both from higher perturbative orders and from the underlying showers.
  Nevertheless, given the behaviour of the fixed-order results, the (N)LO+PS
  predictions appear to be reasonably close to each other and, in addition, by
  including the information on the NLO matrix elements the differences seen at
  LO+PS are reduced.  In particular, it is reassuring that the agreement
  between \herwig and \pythia improves to about 15\% in the medium- and
  high-mass regions.  There is also an improvement at low jet masses (not
  visible in \fig{MJb}), which however should not be over-interpreted, since
  non-perturbative effects are expected to be substantial in this region.
\item The solid-red and dashed-magenta lines in the second inset indicate that
  the \ms results roughly follow similar shape patterns as those of LO+PS
  $W^+bj$, namely that they are softer than NLO+PS $W^+bj$ for \herwig, and
  flat for \pythia. However, in absolute value they are in much better
  agreement than the LO+PS results with the NLO+PS $W^+bj$ predictions.  The
  \ms results show no dependence on the \bwcut~parameter, indicating that
  off-shell effects do not have an impact on the shape of $M(J_b)$.
\item The third inset shows that \herwig and \pythia have a vastly different
  dependence on $\xcut$.  The results showered with \herwig are mostly
  independent of $\xcut$ (except in the low-mass region), while those showered
  with \pythia are highly sensitive to this parameter over the whole range
  considered.  As for the case of $M(W^+,J_b)$, this behaviour can be
  attributed to the different showering models.  The \pythia\ sensitivity of
  the $b$-jet kinematics on $\xcut$ does indeed largely drive the $\xcut$
  variation of the reconstructed-top mass that we have observed
  previously\footnote{We remark that the $\xcut$ dependence of $M(J_b)$ is
    subdominant as far as the behaviour of the reconstructed top mass is
    concerned. In the case of $M(W^+,J_b)$, the $\xcut$ dependence is chiefly
    driven by changes in the direction of flight of the $b$-jet, which renders
    the $M(J_b)-M(W^+,J_b)$ correlation a rather non-trivial function of
    $\xcut$ and of the mass ranges considered.}.  Radiation from the $b$-quark
  is the primary source of mass increase in the region of moderate $b$-jet 
  masses, i.e.~$M(J_b)\gtrsim 10$ GeV, with other phenomena playing
  only a subdominant role there (contributions from splash-in radiation off
  other legs become more important at larger masses, which we do not show in
  \fig{MJb}).  For a vanishing $\xcut$ value, as discussed in item 5 of
  \subsec{mwb}, this radiation is always in direct competition with radiation
  off all other initial and final state partons, forming a single evolution
  chain. Conversely, for non-null values of $\xcut$ the radiation off the $b$
  starts without competition. Thus, it turns out that in this case
  non-competing radiation off the $b$ quark fills the $b$-jet more
  substantially, and leads to a heavier jet. Such an $\xcut$ dependence
  diminishes at larger $M(J_b)$, owing to off-$b$ radiation no longer being
  dominant.  The behaviour at low $M(J_b)\lesssim 10$ GeV is also due to the
  choice of cuts on the $b$-jet. Lowering the $\pt(J_b)$-cut in particular
  leads to a less marked shape difference between the low- and high-$\xcut$
  results.  Crucially, we have checked (with LO simulations) that the same
  $\xcut$-variation pattern in \pythia\ is found when using the local
  recoil. Hence, the striking feature seen in the comparison of the
  \pythia\ and \herwig\ results is not due to the recoil scheme we have
  adopted.
\end{enumerate}
We conclude this section by re-iterating the message that
\fig{MJb} must be seen in its entirety: differences between
generators are as important as perturbative uncertainties,
and can suggest strategies to improve the description of $M(J_b)$.
Furthermore, we point out that this observable is also quite
sensitive to effects controlled by tuning and underlying-event
modelling, which we have not studied here.

\subsubsection{Relative transverse momentum of primary $b$-jet, 
$p_{T,rel}(J_b)$\label{subsec:ptrel}}
%%%%%%%%%%%%%%%%%%%%%%%%%%%%%%%%%%%%%%%%%%%%%%%%%%%%%%%%%%%%%%%%%%%
\begin{figure}[h]
\hspace{-0.6cm}
\includegraphics[trim=0.0cm 0.1cm 0.0cm 0.1cm,clip,width=1.10\textwidth]{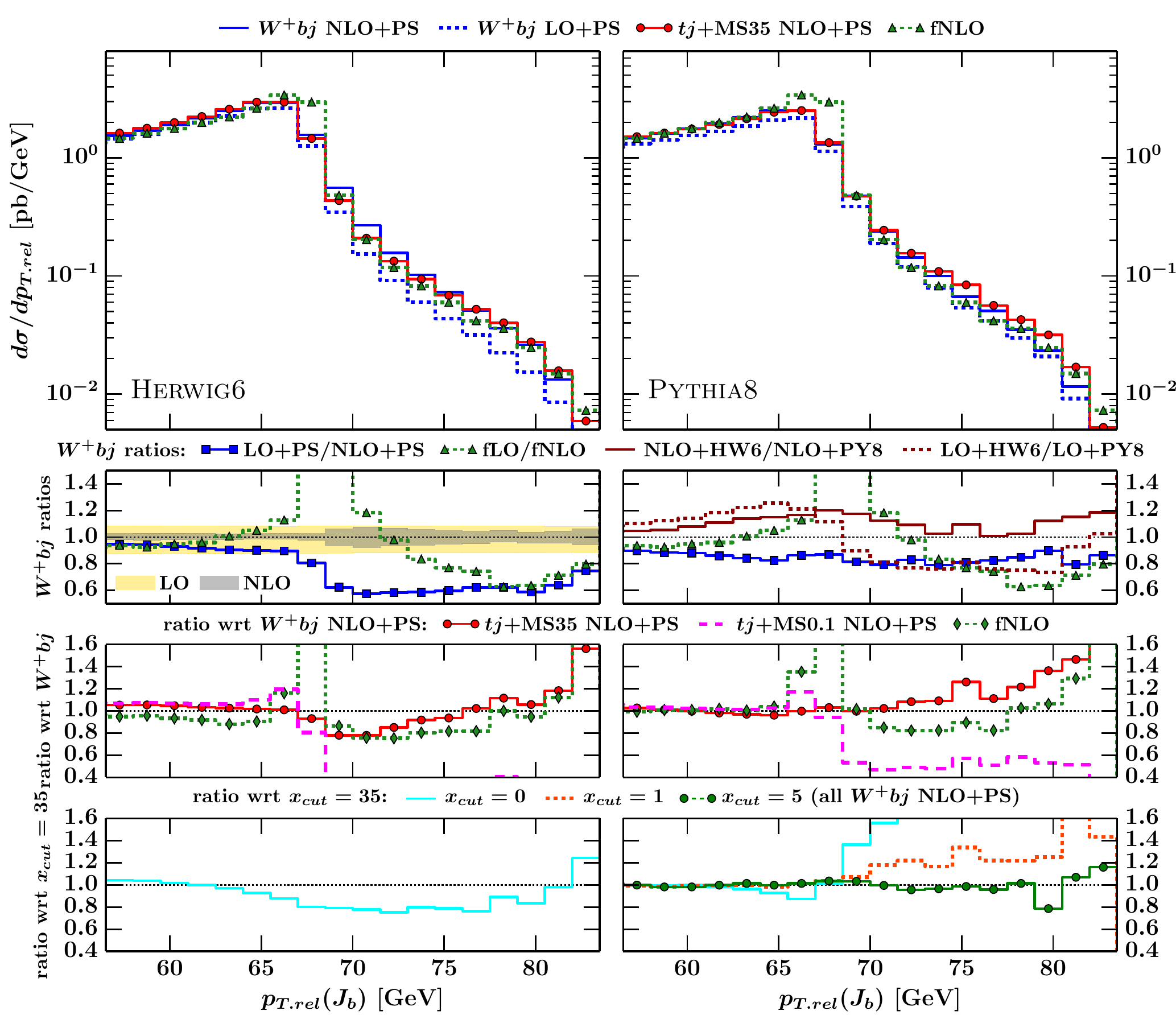} 
\caption{Transverse momentum of primary $b$-jet, $\pt(J_b)$ in top quark rest frame, relative to the direction of flight of the reconstructed top quark.}
\label{fig:pTrelJb}
\end{figure}
%%%%%%%%%%%%%%%%%%%%%%%%%%%%%%%%%%%%%%%%%%%%%%%%%%%%%%%%%%%%%%%%%%%

\noindent
We now turn our attention to the transverse momentum of the primary
$b$-jet in the reconstructed top quark rest frame, relative to the direction
of flight of the reconstructed top quark. We denote this quantity by 
$p_{T,rel}(J_b)$, and display it in
\fig{pTrelJb}.  This observable is challenging to simulate accurately because
its shape is -- already at fixed order -- the result of a balance between
different kinematical effects.
The sharp edge present in this distribution corresponds to the value
$p_{T,rel}(J_b)=(m_t^2-m_W^2)/2m_t$. In the NWA at fLO, transverse momenta
larger than this threshold are kinematically forbidden; the tail beyond the
edge starts appearing at fNLO, due to real corrections to the production
subprocess.  
NLO corrections to the decay become important near the peak
of the distribution, whilst at the peak and above it becomes crucial to treat
the top quark as off-shell \cite{Papanastasiou:2013dta}.  The shoulder of the
distribution and the region above the peak are shaped by an interplay of
contributions from real emissions that originate from the production
subprocess, as well as resonant/non-resonant interference effects and pure
non-resonant effects -- the latter increasing in importance the further one
goes into this region of phase space.
The sensitivity of the shoulder and the tail of this distribution
to $m_t$ make it potentially a good observable for $m_t$-extraction, provided
that the theoretical systematics are under control. Additionally, observables
such as this one are well-suited to disentangling top signals from QCD
backgrounds.  Therefore, it is imperative that MC predictions are fully
understood, and faithfully describe any significant effects over the full
range of $\pt$.  

\begin{enumerate}\itemsep3pt \parskip0pt \parsep0pt
\item Overall, it is apparent that the fixed-order $K$-factor, as in the case
  of $M(W^+,J_b)$, is smoothed out by the showers.  In the region of the
  shoulder and below the peak, LO+PS and NLO+PS predictions are similar in
  shape for both showers.  However, at the kinematic threshold and beyond, the
  NLO corrections result in a ``step up'' in \herwig whilst in \pythia they
  do not induce such a change in shape. This is an indication of how copiously
  the \pythia shower populates this region with radiation already at LO.
\item The effects of parton showering are mild in the low-$p_{T,rel}(J_b)$
  region (especially for \pythia), while they are very large at the sharp edge
  and also visible in the high-$p_{T,rel}(J_b)$ tail.  The sharp edge at
  fixed order is made less steep through the combination of two effects.
  Firstly, near the edge (i.e. close to the fixed-order kinematical threshold)
  multiple FSR emissions off the $b$-quark leaking out of the $b$-jet lead to
  the lowering of the peak.  Secondly, emissions from the production process
  captured inside $J_b$ enhance the region beyond the sharp edge.  We note
  that the high-$p_{T,rel}(J_b)$ region is predominantly LO-accurate in the
  context of the simulations performed in this paper\footnote{This is strictly
    true in the NWA, while off-shell effects partially fill the region beyond
    threshold already at fLO in the full $W^+bj$ computation.}, hence the
  results are expected to have a larger sensitivity to the various
  approximations here than elsewhere. This is reflected in the shape of the
  scale-uncertainty band in the top-left inset.
\item Considering the balance of different effects resulting in the shape at
  fixed order, the relative agreement between \pythia and \herwig for this
  observable is encouraging.  There is a clear improvement in the agreement
  when passing from LO+PS to NLO+PS, even in the high-tail region, where the
  cross section is reduced by 2--3 orders of magnitude with respect to the
  peak.  The pattern of the \herwig\ over \pythia ratio at the LO+PS level can
  be understood as due to the larger amount of radiation (from production)
  in \pythia, which enhances the region above threshold, and thus, by
  unitarity, decreases the cross section below threshold.  At NLO+PS this
  feature is very much reduced, with discrepancies smaller than 15\% in most
  bins.
\item For values of $p_{T,rel}(J_b)$ below the sharp edge, all \ms results do
  a very good job in approximating the shape of the full result in both
  showers. However, beyond threshold, the differences between the two \ms
  $\bwcut$ values, as well as the differences of these w.r.t.~the full result,
  begin to grow significantly, reaching 50\% or more at the edge of the range
  considered. By moving deeper into this region of phase space, as also pointed 
  out in
  \mycite{Papanastasiou:2013dta}, the result becomes increasingly sensitive to
  off-shell and non-resonant effects.  Non-resonant contributions are missing
  in the \ms results and therefore the differences between these and the
  complete NLO+PS $W^+bj$ result can be expected.  This comparison indicates
  that for this observable and in these regions of phase space the full 
  NLO+PS $W^+bj$ computation is a prerequisite for a reliable description.
\item There is a striking difference in the dependence of the \herwig and
  \pythia results on the $\xcut$ parameter.  This is particularly pronounced
  beyond the edge in the spectrum, i.e.~in the region where both real
  radiation and non-resonant effects are very important.  The \herwig
  distribution displays at most a 20\% dependence on $\xcut$.  On the other
  hand, beyond the sharp edge the results showered with \pythia become very
  sensitive to $\xcut$, a feature that stems from the same reasons as those 
  already described in detail in item 5 of \subsecs{mwb}{mjb}.
\end{enumerate}

\subsubsection{Invariant mass of lepton+$b$-jet system, $M(l^+,J_b)$
\label{sec:more-obs}}
The relative transverse momentum discussed in \subsec{ptrel}
can be seen as a member of a class of observables characterised by
the presence of sharp edges due to thresholds in the NWA. Other 
examples are the invariant mass of the $b$-jet-lepton pair, $M(l^+,J_b)$,
and the transverse mass of the system composed of the $b$-jet, the charged
lepton, and the neutrino\footnote{Strictly speaking, the pseudo-top 
mass itself also belongs to this category, but being extremely peculiar
it constitutes a case on its own.}.

These observables may have fairly different properties from the
experimental viewpoint (in particular as far as the reconstruction
of the candidate top pseudo-particle is concerned), the discussion
of which is beyond the scope of the present work. On the other
hand, because of the kinematic features they have in common,
they display similar patterns in terms of the various theoretical
approximations that can be used to predict them. 

\begin{figure}[h]
\hspace{-0.6cm}
\includegraphics[trim=0.0cm 0.1cm 0.0cm 0.1cm,clip,width=1.10\textwidth]{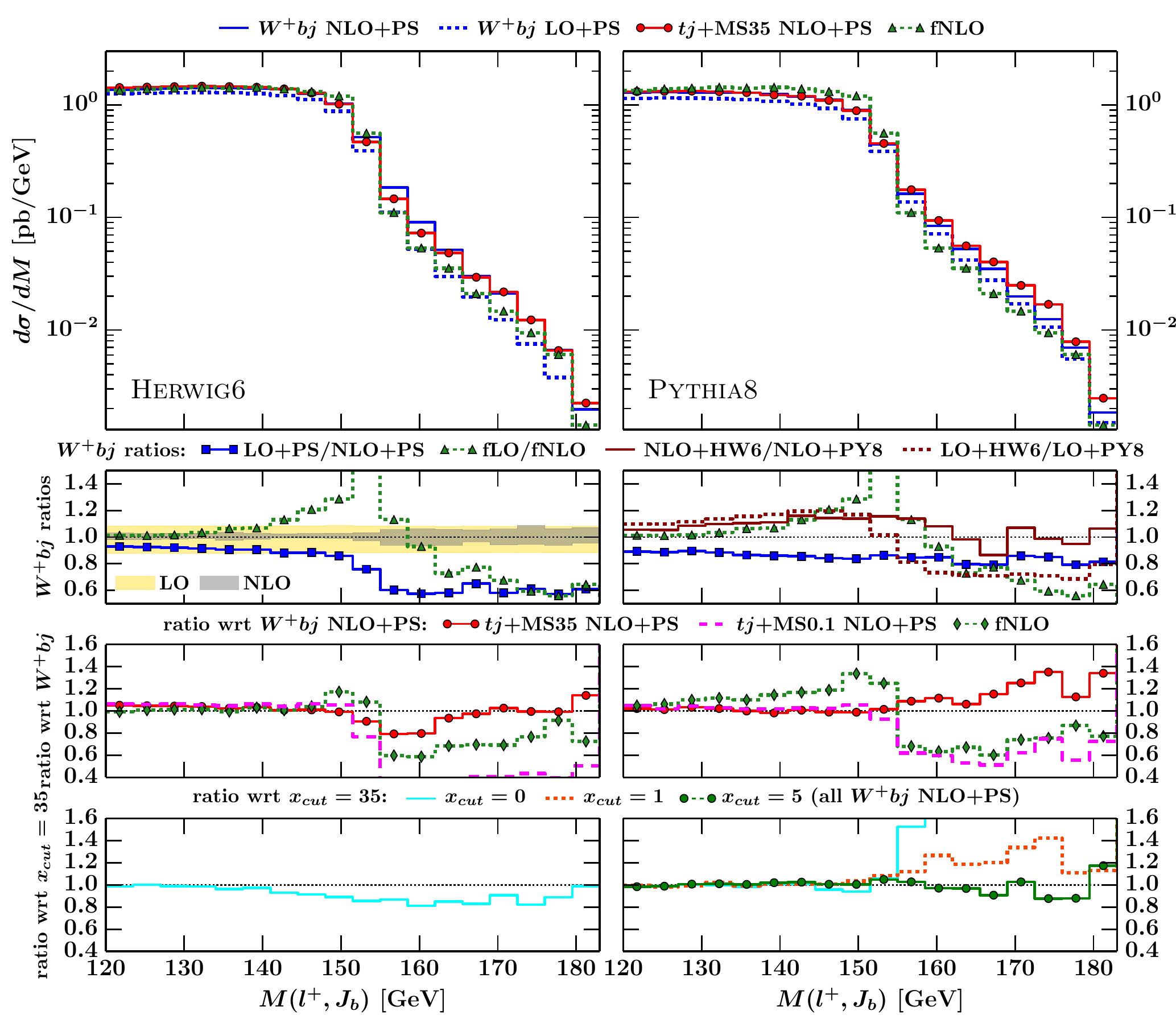} 
\caption{Invariant mass of lepton+$b$-jet system, $M(l^+,J_b)$.}
\label{fig:MJbl}
\end{figure}

To exemplify this fact, in \fig{MJbl}, we show the results for 
$M(l^+,J_b)$. This observable
is often used as a discriminating variable to help disentangle 
top-quark signal events from backgrounds, and interestingly it has
been employed to extract the top mass in $t$-channel-enhanced 
events~\cite{st-tch-mass}. As can be seen from the plots of
\fig{MJbl}, the conclusions of \subsec{ptrel} also apply, largely
unchanged, to the present case.

%%%%%%%%%%%%%%%%%%%%%%%%%%%%%%%%%%%%%%%%%%%%%%%%%%%%%%%%%%%%%%%%%%%
%%%%%%%%%%%%%%%%%%%%%%%%%%%%%%%%%%%%%%%%%%%%%%%%%%%%%%%%%%%%%%%%%%%
Given that the different predictions have been studied in detail and are
now available, it would be interesting to exploit these to quantify the
systematic error on the extraction of $m_t$ via $M(l^+,J_b)$ that is
introduced through the use of generators that do not include full 
off-shell effects and NLO top-decay corrections.

\section{Conclusions and outlook\label{sec:conclusion}}
In this paper we have studied $t$-channel single-top hadroproduction,
where full off-shell and non-resonant effects are computed at the 
matrix-element level with NLO accuracy in QCD, and matching to
parton showers is included. We have done so in the context of
the automated \mga\ framework, where we have implemented our solutions
in a process-independent way. This constitutes the first example,
in the \mcnlo\ approach, of the matching of an NLO calculation 
to a parton shower that features an improved treatment of intermediate 
coloured resonances.

We have considered two Monte Carlos, \pythia\ and \herwig, as 
representatives of different behaviours with regard to several
characteristics, most notably shower evolution and handling of resonances.
This has allowed us to deal in detail with a few resonance-specific
aspects of the matching. Among these were the definition of the MC 
counterterms 
necessary in the \mcnlo\ formalism, the treatment of resolved MC emissions 
off intermediate top quarks, and the writing of information on these tops
in Les-Houches event files. The latter item, to a large extent, pertains
to MC modelling, and it is thus important to keep in mind that the 
writing of an intermediate top in the hard-event file can significantly 
affect how the events look after parton showering. Although we have
argued that including such information is certainly physically motivated,
and is consistent with results obtained in the $\Gamma_t\to 0$ limit,
we have studied its consequences by parametrising it by means of an
arbitrary dimensionless quantity $\xcut$. It will be interesting to 
compare theoretical predictions, and their dependence on $\xcut$,
with actual data.

We have obtained results based on a generic final-state analysis, chosen to be
as similar as possible to that of the fixed-order calculation presented in
\mycite{Papanastasiou:2013dta}, in order to allow for a direct comparison to
the latter paper.  Overall, it is observed that for this typical analysis the
differential $K$-factors at the hadron level can be large and non-constant in
shape. However, as expected, the scale dependence of the results generally
decreases when going to NLO+PS accuracy.  Our comparison to the fNLO
predictions illustrates that effects of parton showering and hadronisation can
be large ($>$10-15\%).  We also find that the agreement between the
predictions of the two showers improves at NLO, though the differences
themselves are found to be sizeable for some observables (in particular for
the invariant mass of the primary $b$-jet).

An important consequence of the availability of predictions at NLO+PS
accuracy, with full off-shell effects, is that existing approximations can be
scrutinised and validated. We have made a detailed comparison to the results
of stable single-top production at NLO+PS accuracy, where the spin-correlated
decay of the top quark and the leading off-shell effects are included by 
using \ms. On the whole, it is observed that the \ms results describe the full 
results remarkably well, which is certainly encouraging
in view of the fact that the predictions of the former type presently 
constitute the benchmark for simulations that involve coloured resonances
at LHC experiments. However, we have observed some notable differences
between stable-top and full results in some of the distributions we have
studied, with an observable-dependent pattern. We have attributed these to 
NLO effects in the top-quark decay and to non-resonant effects, neither of
which can be described by the \ms procedure, and to off-shell effects,
which \ms\ can simulate only in an approximate manner. We thus conclude
that, when targeting a less-than-10\% accuracy, a better description of
intermediate resonances is a necessity.

There are several aspects of this paper that will be interesting to
consider in future work. The most obvious is the application of
our findings to the matching of $t\bar{t}$ production with full
off-shell effects ($W^+W^-b\bar{b}$) to parton showers. Furthermore,
in view of the sometimes large differences between the \pythia\ and
\herwig\ predictions, it will also be worthwhile performing a careful
study of the impact of shower initial conditions. This is because single-top
production is an example of a multi-scale process for which a more
sophisticated choice of shower scales might be required for an
improved phenomenology treatment (as recently observed e.g.~in
ref.~\cite{Wiesemann:2014ioa}, which also deals with the presence of 
final-state $b$ quarks).
Finally, we point out that our results matched to the \pythia shower show
qualitative agreement with those presented in \mycite{Jezo:2015aia}, which
employ the {\sc POWHEG} matching scheme.  A thorough comparison of the two
approaches using the same inputs and analysis setup would obviously be of
great interest, and one which we intend to pursue.

\section*{Acknowledgements}
We would like to thank Bryan Webber for many useful and illuminating
discussions on this topic as well as for providing us with a modified version
of \herwig that allowed the vetoing of PS emissions off intermediate top
quarks.

The work of RF is supported by the Alexander von Humboldt Foundation, in the
framework of the Sofja Kovaleskaja Award Project ``Event Simulation for the
Large Hadron Collider at High Precision," endowed by the German Federal
Ministry of Education and Research.  SF is grateful to CERN TH division for
hospitality during the course of this work.  The work of AP is supported by
the UK Science and Technology Facilities Council [grant ST/L002760/1].  SP is
supported by the US Department of Energy under contract DE-AC02-76SF00515.
The work of PT has received funding from the European Union Seventh Framework
programme for research and innovation under the Marie Curie grant agreement
N. 609402-–2020 researchers: Train to Move (T2M).  This work has been supported
in part by the ERC grant 291377 ``LHCtheory: Theoretical predictions and
analyses of LHC physics: advancing the precision frontier".

%%\newpage
\appendix

\section{Technicalities on the treatment of resonances\label{app:reso}}
In this appendix we sketch the implementation in \mga of the
type-IIb solution alluded to in \sec{genFKS}, for the
case of final-state singularities relevant to the FKS sector where 
the FKS pair belongs to the tree whose root is the resonance $\beta$.
In order to simplify the discussion as much as possible, we shall proceed
as in \sec{gen}, i.e.~pretending that only soft singularities
are present. However, as mentioned before, in \mga all counterevents associated 
with a given event have the same reduced kinematics. Therefore, the actual
formulae implemented in the program differ from those given below only by
marginal technical aspects.
We also recall that in the case of the singularities we are interested in,
the phase-space parametrisation employed by \mga is that of section~5.2 of 
\mycite{Frixione:2007vw}, and its generalisation to the case of a massive 
FKS sister. This implies that, using the labelling of \fig{diag1}:
\beq
\kbt(\phivar,\xi)=\left(k_\gamma+k_\mu\right)^2\,,\;\;\;\;\;\;\;\;
\kbt(\phivar,0)=\left(\bar{k}_\gamma+\bar{k}_\mu\right)^2\,,
\label{eq:FNO1}
\eeq
where the barred four-momenta $\bar{k}$ conventionally denote those
of the counterevent configuration associated with the event configuration
whose four-momenta are denoted by un-barred symbols $k$. Furthermore
\beq
\bar{k}_\gamma=\boost\, k_\gamma\,,\;\;\;\;\;\;\;\;
\bar{k}_\mu\ne\boost\, k_\mu\,,
\label{eq:FNO2}
\eeq
with $\boost$ a boost. Therefore, \eq{kbcond} is not fulfilled,
whence the necessity of a type-II solution.

In order to proceed, let us start from the basic expression of the
subtracted cross section, \eq{sig1}, that we can replace with:
\beq
\int db\int_0^{\ximax(b)} d\xi \frac{1}{\xi}\left[
\sigma\!\left(\kbt(b,\xi)\sep b,\xi\right)-
\sigma\!\left(\kbt(b,0)\sep b,0\right)\right]\,,
\label{eq:sig4}
\eeq
where:
\beq
\varsigma(b,\xi)=\sigma(b,\xi)\stepf(\ximax(b)-\xi)\,.
\label{eq:varSdef}
\eeq
Equation~(\ref{eq:sig4}) differs from \eq{sig1} 
by a contribution due to the integral of the counterterm in the range
$\xi>\ximax(b)$ (owing to the fact that for the soft counterterm the $\stepf$
function of \eq{varSdef} is identically equal to one). As already
discussed in \sec{gen}, Born-like terms do not pose significant
problems in the presence of resonances. We shall thus ignore this 
contribution in what follows, and deal solely with \eq{sig4}. In \mga,
the parametrisation of Born-level integration variables in the context
of an NLO computation is identical to that adopted for a tree-level
computation of the same multiplicity (see \mycite{Frederix:2009yq}).
This implies that, by construction, one of the variables $b$ will
coincide with the virtuality of the resonance $\beta$, computed with
the counterevent kinematics. Let us denote this variable by $\bbe$,
and all of the other integration variables collectively by $\bnotbe$:
\beq
b=\{\bbe,\bnotbe\}\,.
\label{eq:variables}
\eeq
By construction, one has:
\beq
\bbe=\kbt(b,0)\equiv \kbt(\bbe,\bnotbe,0)\,,
\label{eq:bbe}
\eeq
whereas, owing to \eqs{FNO1}{FNO2}:
\beq
\bbe\ne\kbt(\bbe,\bnotbe,\xi)\,.
\label{eq:bbeEV}
\eeq
Equation~(\ref{eq:bbeEV}) suggests that 
$\kbt(\bbe,\bnotbe,\xi)$, seen as a function of $\bbe$ at fixed
$(\bnotbe,\xi)$, can be identified with $\Phi_\xi$ of \eq{phspmap}
for a one-dimensional change of integration variables for a 
type-IIa solution; its inverse can be used for a type-IIb solution.

We have considered both types of approaches, and found that the type-IIa
one did not perform in a satisfactory manner from the numerical viewpoint.
The reason is the following: with a type-IIa solution,
both event and (re-mapped) counterevents 
have a Breit-Wigner peak at $\kbt(\bbe,\bnotbe,\xi)\simeq\mbt$.
This implies that the integration variable $\bbe$ will {\em not}
be peaked at $\mbt$, but at a somewhat different (typically lower)
value. Moreover, and more importantly, the position of such a peak
will be correlated with the value of $\xi$ (and, in the actual QCD case
where collinear singularities are present, with that of the
angle between the FKS parton and its sister). The first issue
renders it difficult to guess analytically an efficient change of variables 
from the relevant ``Vegas $x$'' to $\bbe$, which implies a longer-than-desired 
grid optimisation, while the second issue effectively hampers such an 
optimisation (since correlations are notoriously difficult to handle 
in adaptive integrations). For this reason, our solution of choice
in \mga is a type-IIb one, which we now proceed to describe in
greater detail.

As discussed in \sec{gen}, type-IIb solutions entail
the manipulation of the event contribution. We single out such
contribution in \eq{sig4}, which we implicitly and temporarily
regularise (e.g.~with a cutoff) in order to avoid divergences, 
and rewrite it as follows:
\beqn
E&=&\int db^\prime\int_0^{\ximax(b^\prime)} \frac{d\xi}{\xi}
\sigma\!\left(\kbt(b^\prime,\xi)\sep b^\prime,\xi\right)
\nonumber
\\*&\equiv&
\int d\bbe^\prime d\bnotbe^\prime\int_0^1 \frac{d\xi}{\xi}
\sigma\!\left(\kbt(\bbe^\prime,\bnotbe^\prime,\xi)\sep 
\bbe^\prime,\bnotbe^\prime,\xi\right)\,
\stepf\!\left({\ximax(\bbe^\prime,\bnotbe^\prime)}-\xi\right)\,,
\phantom{aaa}
\label{eq:sigev}
\eeqn
having trivially renamed $b$ as $b^\prime$. We then perform the
following change of integration variables:
\beq
\{\bbe^\prime,\bnotbe^\prime\}\longrightarrow \{\bbe,\bnotbe\}\,,
\;\;\;\;\;\;
\bnotbe^\prime=\bnotbe\,,
\;\;\;\;\;\;
\bbe^\prime=\kbtmo(\bbe,\bnotbe,\xi)\,,
\label{eq:change3}
\eeq
whence \eq{sigev} becomes:
\beqn
E&=&\int d\bbe d\bnotbe\int_0^1\frac{d\xi}{\xi}\,
\frac{\partial\kbtmo(\bbe,\bnotbe,\xi)}{\partial\bbe}\,
\stepf\left({\ximax(\kbtmo(\bbe,\bnotbe,\xi),\bnotbe)}-\xi\right) 
\nonumber
\\*&&\phantom{\int d\bbe d\bnotbe\int_0^1\frac{d\xi}{\xi}\,}\times
\sigma\!\left(\bbe\sep\kbtmo(\bbe,\bnotbe,\xi),\bnotbe,\xi\right)\,.
\label{eq:sigev2}
\eeqn
The first argument of $\sigma$ in \eq{sigev2} shows that the event now has 
the desired property (thanks to \eq{bbe}), namely that the reconstructed 
invariant mass of the resonance is equal to that of the counterevent 
(generated with the same $\{\bbe,\bnotbe\}$ and not re-mapped), in keeping 
with the general derivation of type-IIb solutions. A drawback of \eq{sigev2}
is the possible difficulty of computing the jacobian analytically; in
\mga we bypassed this problem by resorting to entirely numerical
methods, with excellent performances in terms of stability and accuracy.
The re-mapped subtracted cross section can finally be obtained by replacing
the event contribution to \eq{sig4} with the r.h.s.~of
\eq{sigev2}. Such a form is essentially what is implemented
in \mga, barring numerically-small contributions due to the following
features.
\begin{itemize}
\item The support of $\kbt(\bbe,\bnotbe,\xi)$ is in general different
w.r.t.~that of its inverse. In this case, where either the event
or the counterevents are equal to zero, no re-mapping is performed.
\item It may happen that, for certain values of $\xi$ (typically far from 
zero, and with a massive FKS sister) and close to the borders of the support 
of $\kbt(\bbe,\bnotbe,\xi)$, such a function is not monotonic. Although
one could carry out the procedure outlined above in a piece-wise manner, 
we have opted for not performing the re-mapping in such a case.
\end{itemize}
Finally, we point out that in our code the integration variable
relevant to the FKS soft subtraction is not $\xi$, but actually
its rescaled version $\xih$, defined so that $\xi=\xih\ximax(b)$.
This guarantees a better numerical performance, chiefly owing 
to the complete absence of correlations between $\xih$ and
other integration variables.

%%\newpage
\phantomsection
\addcontentsline{toc}{section}{References}
\bibliographystyle{JHEP}
\bibliography{wbj_ps_refs.bib}

\end{document}